\documentclass[12pt]{article}
\usepackage{amssymb,amsmath}
\usepackage{latexsym}
\usepackage{graphicx}
\usepackage[linktocpage=true]{hyperref}
\makeatletter
\@addtoreset{equation}{section}
\makeatother

\renewcommand{\theequation}{\arabic{section}.\arabic{equation}}

\def\coeff#1#2{\relax{\textstyle {#1 \over #2}}\displaystyle}

\def\IR{\mathbb{R}}
\def\ZZ{\mathbb{Z}}

\def\ds{\displaystyle}
\newcommand{\be}{\begin{equation}}
\newcommand{\ee}{\end{equation}}
\newcommand{\bea}{\begin{eqnarray}}
\newcommand{\eea}{\end{eqnarray}}

\begin{document}

\begin{titlepage}

\begin{flushright}
IPhT-T10/081
\end{flushright}

\bigskip
\bigskip
\centerline{\Large \bf An Infinite-Dimensional Family of }
\smallskip
\centerline{\Large \bf Black-Hole Microstate Geometries}
\medskip
\bigskip
\bigskip
\centerline{{\bf Iosif Bena$^1$, Nikolay Bobev$^2$, Stefano Giusto$^3$,}}
\centerline{{\bf Cl\'{e}ment Ruef$^{\, 1}$ and Nicholas P. Warner$^2$}}
\bigskip
\centerline{$^1$ Institut de Physique Th\'eorique, }
\centerline{CEA Saclay, CNRS-URA 2306, 91191 Gif sur Yvette, France}
\bigskip
\centerline{$^2$ Department of Physics and Astronomy}
\centerline{University of Southern California} \centerline{Los
Angeles, CA 90089, USA}
\bigskip
\centerline{$^3$ Dipartimento di Fisica}
\centerline{Universit\`a di Genova} \centerline{via Dodecaneso, 33, I-16146, Genoa, Italy}
\bigskip
\centerline{{\rm iosif.bena@cea.fr,~bobev@usc.edu,~giusto.stefano@ge.infn.it, } }
\centerline{{\rm clement.ruef@cea.fr,~warner@usc.edu} }
\bigskip
\bigskip

\begin{abstract}

We construct the first explicit, smooth, horizonless black-hole microstate geometry whose moduli space is described by an arbitrary function of one variable and is thus infinite-dimensional. This is achieved by constructing the scalar Green function on a simple D6-${\rm \overline{D6}}$ background, and using this Green function to obtain the fully back-reacted solution for a supertube with varying charge density in this background. We show that this supertube can store parametrically more entropy than in flat space, confirming the entropy enhancement mechanism that was predicted using brane probes.   We also show that all the local properties of the fully back-reacted solution can, in fact, be obtained using the DBI action of an appropriate brane probe. In particular, the supergravity and the DBI analysis yield identical {\it functional} bubble equations that govern the relative locations of the centers. This indicates that there is a non-renormalization theorem that protects these functional equations as one moves in moduli space. Our construction creates configurations that are beyond the scope of recent arguments that appear to put strong limits on the entropy that can be found in smooth supergravity solutions.

\end{abstract}

\end{titlepage}
%%%%%%%%%%%%%%%%%%%%%%%%%%%%%%%%%%%%%

\tableofcontents
%%%%%%%%%%%%%%%%%%%%%%%%%%%%%%%%%%%%%
\newpage
%%%%%%%%%%%%%%%%%%%%%%%%%%%%%%%%%%%%%
\section{Introduction}
%%%%%%%%%%%%%%%%%%%%%%%%%%%%%%%%%%%%%

One of the key aspects to understanding black holes within string theory involves 
determining the families of supergravity solutions that have a given set of charges and angular
momenta, and examining the relationship between the number of these
solutions and the Bekenstein-Hawking entropy of a black hole with
identical charges and angular momenta.

The interest in this question is spurred by the desire to establish if, and in what form, the so-called fuzzball proposal (see
\cite{Mathur:2005zp,Bena:2007kg,Skenderis:2008qn,Balasubramanian:2008da,Chowdhury:2010ct}
for reviews) applies to BPS black holes. This proposal requires that, within string theory, there should exist a huge number of horizonless configurations with unitary scattering, that have the same mass, charges and angular momenta as a black hole, and that start differing from each other at the location of the would-be black hole horizon. Furthermore, the number of such configurations should reproduce the Bekenstein-Hawking entropy of the black hole. If enough such configurations exist, then the $AdS$-CFT correspondence would strongly suggest that the classical black hole solution only gives a thermodynamic approximation of the physics that is correct at scales larger than the horizon scale, but does not correctly describe the physics at the horizon, much as the thermodynamic description of a gas breaks down at scales smaller than the mean free path.

The construction and investigation of black hole microstate configurations that has been going on over the past few years has yielded several interesting pieces of information.  First, there is a very interesting geometric transition that takes the original black-hole or black-ring geometry over to a huge moduli space of smooth, horizonless bubbled geometries  \cite{Bena:2005va,Berglund:2005vb, Bena:2007kg}.  Amongst such bubbled geometries  there exist a very large number of smooth horizonless solutions that have the same charges and angular momenta as a BPS black hole with a macroscopically-large horizon area, both in five and in four dimensions \cite{Bena:2006is,Bena:2007qc,Balasubramanian:2006gi}. Furthermore, these solutions can have a very long throat, and one can argue \cite{Bena:2006is,deBoer:2008zn} that they should be dual to CFT states that live in the same CFT ``typical'' sector as the states that give rise to the Bekenstein-Hawking entropy of the black hole.

For practical, computational reasons it is easiest to study the microstate solutions that have a Gibbons-Hawking (GH) base space and thus descend to multi-center solutions in four dimensions \cite{Bates:2003vx,Denef:2002ru}.  There are very large families of such solutions, however the semi-classical quantization of the moduli space shows that these solutions give an entropy that is parametrically smaller than the entropy of a black hole with the same charges \cite{deBoer:2008zn,deBoer:2009un}. This is to be expected, and indeed was anticipated in the earlier, very limited estimates of \cite{Bena:2006is}:  One cannot hope that counting only microstates that have a large amount of symmetry could reproduce the entropy of a given system. The simple bubbled solutions in five dimensions have the symmetries of the compactification  torus as well as a tri-holomorphic $U(1)$ isometry along the Gibbons-Hawking (GH) fiber. From such a class of solutions one cannot even reproduce the entropy of the two-charge D1-D5 system, which is known to come from smooth horizonless solutions \cite{Lunin:2001fv,Maldacena:2000dr,Lunin:2002iz}. The entropy of supergravity $U(1)$-invariant solutions was also matched against the entropy of a weakly coupled graviton gas in $AdS_3 \times S^2$ in  
\cite{deBoer:2009un}. While this entropy is not restricted to $U(1)$-invariant configurations, it is intrinsically perturbative, and cannot capture the supertubes we consider here; in the D1-D5-P frame our configurations give rise to a non-trivial topology, and in other frames have dipole charges corresponding to non-perturbative brane configurations. 

To some extent, the amazing part of $U(1)$-invariant bubbling geometries and of four-dimensional multi-center solutions is not the fact that there are so few microstates with a set of isometries (in the D1-D5 system there is only one such microstate), but that there are {\it so many}. The importance of the bubbled GH geometries is not so much in the large number of highly symmetric solutions  but in the fact that they come from a geometric transition of branes in flat space, and this transition opens up vast new families of solutions in which the bubbles of these geometries change shape, or ``wiggle,'' in ways that break the original isometries 
and depend upon internal dimensions that were previously frozen out of the dynamics. It is one of the primary purposes of this paper to find classes of such wiggling solutions that depend upon arbitrary functions.

We therefore expect that, in the vicinity of the highly symmetric solutions counted in \cite{deBoer:2008zn,deBoer:2009un},  there will be many smooth solutions that do not preserve the isometries.  Indeed, the simplest class of such solutions can be obtained by putting BPS supertubes of arbitrary shape \cite{Mateos:2001qs} in one of the bubbling microstate solutions \cite{Bena:2008dw}. Through spectral flow one can then think of these solutions as fluctuating bubble geometries \cite{Ford:2006yb,Bena:2008wt}.  These solutions depend on several continuous functions, and hence have an infinite-dimensional moduli space, of which the finite-dimensional moduli space counted in \cite{deBoer:2008zn,deBoer:2009un} is a very small, discrete subset. If one naively counts these fluctuating solutions, ignoring the back-reaction of the
supertubes\footnote{Much in the same spirit as counting the D4-D0  black hole entropy in the regime of parameters where the D0's can be  treated as probes \cite{Gaiotto:2004ij,Denef:2007yt,Gimon:2007mha}.}, the entropy one obtains from such a calculation is infinite
\cite{Bena:2008nh}.  However, if one assumes that the length of the throat in which the supertube sits caps off to the value expected from
the dual CFT to correspond to the typical black hole microstates, the entropy of these supertube can potentially have the same growth with the charges as the entropy of a black hole \cite{Bena:2008nh}.

It is also worth noting that the technique of using wiggling supertubes to create fluctuating bubbled geometries is a convenient mathematical device for creating an explicit, though rather restricted class of such fluctuating geometries.  There are, of course, considerably more general classes of BPS fluctuating geometries, and conceivably there might even be BPS classes that involve fluctuating, regular surfaces with arbitrary functions of two variables or non-geometric backgrounds \cite{deBoer:2010ud}.

Returning to fluctuating geometries created by wiggling supertubes, it would be clearly interesting to use the technology developed in
\cite{deBoer:2008zn,deBoer:2009un}, as well as earlier technology developed in \cite{Crnkovic:1986ex,Grant:2005qc,Rychkov:2005ji} to
count the back-reacted wiggly supertube solutions in its infinite-dimensional moduli space (as was done for two-charge supertubes
in flat space in \cite{Rychkov:2005ji}). Given that the back-reaction of both the supertubes in flat space \cite{Lunin:2001fv,Maldacena:2000dr,Lunin:2002iz} and the ones in more complicated three-charge backgrounds \cite{Bena:2008dw} is expected to lead to smooth supergravity solutions\footnote{These solutions are smooth only in the duality frame in which the supertubes have dipole charge corresponding to KK monopoles, but not in other duality frames \cite{Emparan:2001ux,Lunin:2001fv,Maldacena:2000dr,Lunin:2002iz,Bena:2008dw}.}, if the entropy one finds this way is black-hole-like then this would go a long way toward establishing the validity of the fuzzball proposal for extremal BPS black holes.

We have several goals in writing this paper. We begin by outlining a procedure by which one can construct explicit supergravity solutions that contain wiggly supertubes, and that depend on arbitrary functions.  Even if all such solutions can be formally written down using Green functions \cite{Bena:2004de}, constructing them explicitly is not straightforward because one needs both scalar and vector Green functions in multi-center Taub-NUT. The scalar ones were constructed by Page for a regular multi-center Taub-NUT space  \cite{Page:1979ga}, but the vector ones are not known. An upcoming paper contains the construction of these Green functions, which are much more involved than the scalar ones. Even if one could write implicitly a full wiggly-supertube solution in this way, one cannot integrate explicitly the Green functions around an arbitrarily-shaped supertube profile. We therefore focus on a sub-class of wiggly supertubes where the supertube shape remains round, but the charge densities inside the supertube world-volume fluctuate.  As we will show in Section \ref{sec:solution}, this class of solutions is parametrized by one arbitrary continuous function. Since the supertube shape is unchanged, the dipole magnetic fields of these solutions are exactly the same as those of the solution with a tri-holomorphic $U(1)$ invariance. All of this makes it just possible to construct explicitly the fully back-reacted solution.  It is thus our primary goal here to construct such a fully back-reacted wiggly solution.

Constructing a wiggling supertube with one arbitrary function is obviously far less general than exciting all possible shape (and fermionic) modes. On the other hand, if the combined shape and fermionic modes of supertubes can capture the entropy of a black hole of finite horizon area, the number of arbitrary functions should then represent the central charge of the underlying field theory.  
Hence, one arbitrary function should be sufficient to capture the growth of the entropy of supergravity solutions as a function of the charges. The overall constant of proportionality can then be trivially restored.  Hence the construction of explicit fluctuating geometries that depend on one arbitrary function is extremely important, because it can be used to capture the contribution to the entropy of the most general class of supertube fluctuations. 

A number of important mathematical and physical issues arise out of the construction of fully-back reacted solutions  for wiggly supertubes and these lead to the other important results of this paper.  

First,  if one wants to construct solutions that have the same charges as a black hole with a macroscopically-large horizon area and that have a hope via the entropy enhancement phenomenon to describe the entropy of this black hole, one must use {\it ambipolar} Taub-NUT or Gibbons-Hawking base spaces, whose signature is both $(+,+,+,+)$ and $(-,-,-,-)$. The five-dimensional solutions constructed using such apparently pathological base-spaces were shown in \cite{Giusto:2004kj,Bena:2005va,Berglund:2005vb,Saxena:2005uk} to actually be smooth and horizonless.  However, in order to construct wiggly supertube solutions one needs to know the scalar Green functions on these base spaces.

Since ambipolar base spaces have a very significant apparent pathology, finding the physically-correct Green function is rather subtle. For example, one expects a Green function to diverge precisely when the distance between the source and the field point goes to zero but in an ambipolar space the vanishing of the distance function from a given point in the $(-,-,-,-)$ region defines a codimension-one ``image'' hypersurface in the $(+,+,+,+)$ region.  However, in the fully back-reacted solution this point is at a finite distance from this zero-distance ``image'' hypersurface, and hence the Green function should not diverge there.

There are several ways in which one can try to find the proper Green function. The most obvious is to analytically continue the Green function on a regular Taub-NUT space first constructed by Page in \cite{Page:1979ga}  (which we call from now on the {\it Page Green function}).  As one might expect, such continuations are bedeviled by the ``image surface'' problem and,  as we will show in Section \ref{Greensfunctions}, this procedure gives a Green function that solves the  correct equations but diverges for non-coincident points. One can similarly try to use parts of the Page Green function and recombine them in such a way as to obtain a physical Green function, but the obvious procedures do not work either. The ``best'' Green function that we could construct using this direct method has cusps at non-coincident points, and at first investigation does not appear to be good-enough for our purposes. It would be interesting to see whether a better-behaved Green function for ambipolar spaces can be constructed directly.

The only effective strategy so far for constructing the requisite Green function on ambipolar spaces is to use the fact that the five-dimensional, Lorentzian geometry is regular, construct a Green function in five dimensions and then reduce it to the proper ambipolar Green function by descending to the base in four dimensions.  As we will see in Section \ref{Greensfunctions}, this indirect method has the advantage of yielding the physical Green function, but has the disadvantage of being technically challenging for even the simplest of   ambipolar spaces. We show how to use this method to construct the Green function for an ambipolar Gibbons-Hawking space with harmonic function ${1\over |\vec r + \vec a|} - {1\over |\vec r - \vec a|}$ starting from the five-dimensional Green function on global $AdS_3 \times S^2$.  We also compare this physical Green function to the possible extensions of the Page Green function, and explore the parallels and differences between the various Green functions.  It is to be hoped that there might be a systematic procedure for adapting the  Page Green function to obtain the physical Green function in general, but the complexity of the differences make this a challenging task.  On the other hand, as we will describe below, we will learn from the one explicit Green function that we can construct that almost all of the essential physics can be extracted via careful use of probe solutions.

Having obtained the Green function for the two-center, zero-charge Gibbons Hawking space, in Section \ref{sec:fullsol} we construct the fully-back-reacted supergravity solutions corresponding to wiggly supertubes in that space. It should also be noted that a family of wiggling supertubes carrying three charges was obtained in  \cite{Ford:2006yb} in a flat-space background.  Here, by working in an ambipolar background we find three-charge solutions that form the largest known family of {\it  black-hole} microstate solutions, parameterized by an arbitrary continuous function, and thus having an infinite-dimensional moduli space.  We also analyze the conditions that these solutions must satisfy in order to be smooth and free of closed timelike curves.  

From our explicit solution for the wiggling supertube, we obtain the bubble, or integrability, equations that relate fluctuating charge densities and determine the location of the wiggly supertube. When the supertube does not wiggle, the solution can be reduced to a three-center solution in four dimensions, and the inter-center positions must satisfy some rather simple  bubble equations. If the supertube changes shape, then it generically does not correspond to a point in the $\IR^3$ base of the Gibbons-Hawking space, and hence the regularity conditions will become more complicated than the simple bubble equations. Nevertheless, if the supertube is round and its wiggles come solely from the density modes, its location is still a point in the $\IR^3$ base, and one can still write bubble equations. These equations are {\it functionals} of the density modes of the supertube and we will see how the ambipolar Greens functions simplify at Gibbons Hawking points and thus lead to relatively straightforward bubble equations on ambipolar bases.

Having obtained these ``bubble functional equations'' using supergravity and our new Green function, we investigate, in Section \ref{sec:DBI}, the extent to which one can obtain the same physical data by studying the Born-Infeld action of a supertube in this space, much as one did for round supertubes.    As a probe, the Born-Infeld dynamics of the supertube is sensitive only to the physics in its vicinity, and hence will only yield the bubble equation for the supertube point.   However, one can use the fact that the bubble equations contain only pairwise interactions between various centers to fish out, from the supertube bubble equations, the various parts of the bubble equations for the other points. This procedure actually yields all the bubble functional equations simply from the supertube Born-Infeld action.

Remarkably, the two sets of functional equations agree with each other, despite being obtained by very different procedures: one set is obtained by going through a very technical construction of a physical Green function on an ambipolar space, then  plugging these Green functions in the supergravity equations and demanding no closed timelike curves, while the other set is obtained by extracting and rearranging terms in the Born-Infeld action of a probe wiggly supertube.  The fact that the bubble functional equations obtained in these two different regimes of parameters agree, points to the existence of a non-renormalization theorem similar to that which protects the bubble equations for $U(1)$ invariant multi-center solutions\footnote{We would like to remind the reader that these
equations can be obtained both at weak effective coupling using quiver quantum mechanics and at strong effective coupling using the fully back-reacted supergravity solution, and their form is the same  \cite{Denef:2002ru,Bates:2003vx}.}. This non-renormalization theorem was very useful for finding the symplectic form of and quantizing the $U(1)$-invariant multi-center solution in \cite{deBoer:2008zn,deBoer:2009un}, and we believe it will prove equally useful in quantizing the infinite-dimensional moduli space of the supergravity solutions we are constructing here.

In Section \ref{sec:enhancement} we return to the full supergravity solution to analyze and estimate the entropy  that can be stored in the fully back-reacted supertube. In \cite{Bena:2008nh} this entropy was estimated using the Born-Infeld action of the supertube (using the technology of \cite{Palmer:2004gu,Bak:2004kz}), and was found to be much larger than that of supertubes in flat space. Indeed, the charges that control this entropy are not the electric charges of the supertubes, but some effective charges that are the sum of the electric charges of the supertube and of a contribution coming from the magnetic fluxes of the background. Hence, even a small supertube, when placed in a background with large magnetic fields can have a considerable entropy. The analysis in \cite{Bena:2008nh} involved probes and so did not include the back-reaction and it was found that the entropy of a supertube placed in scaling solution can grow arbitrarily
large with increasing the depth of the scaling solution. The technology developed in this paper allows us to construct a
fully-back-reacted example of this entropy enhancement mechanism, and to estimate exactly how much entropy can a supertube carry. We find that this entropy is to leading order independent on the electric charges of the supertube, and only depends on its magnetic dipole charge and on the charges and fluxes of the background. This shows that entropy enhancement is indeed a real physical feature of the complete microstate geometry. 

To summarize, in Section \ref{sec:solution} we present our methodology for obtaining fully-back-reacted solutions that depend on one arbitrary continuous function. In Section \ref{Greensfunctions} we explore several possibilities for constructing Green functions for ambipolar spaces, and find the physical Green function for an ambipolar Gibbons-Hawking space with two centers of opposite charge. In Section \ref{sec:fullsol} we use this Green function to construct the fully back-reacted solution corresponding to wiggly supertubes in this space,and find the functional bubble equations satisfied by the positions of the supertube and GH centers. In Section \ref{sec:DBI} we present a way to extract these equations from the Born-Infeld action of the wiggly supertube, and find that the two sets of functional equations agree. In Section \ref{sec:enhancement} we use our results to show that the entropy of a supertube inside a bubbling solution depends on the charges of the background and not on its explicit charges, and thus establish the existence of the entropy
enhancement mechanism proposed in \cite{Bena:2008nh}. We conclude in Section 7. Appendix A contains some more details of the Green
function constructed in Section \ref{Greensfunctions}.

%%%%%%%%%%%%%%%%%%%%%%%%%%%%%%%%%%%%%
\section{The general back-reacted solution with fluctuating electric charge densities}
\label{sec:solution}
%%%%%%%%%%%%%%%%%%%%%%%%%%%%%%%%%%%%%

Three-charge solutions with four supercharges are most simply
written in the M-theory duality frame in which the three charges
are treated most symmetrically and correspond to three types of M2
branes wrapping three $T^2$'s inside $T^6$ \cite{Bena:2004de,Gutowski:2004yv}. The
metric is:
\begin{multline}
ds_{11}^2  = - \left( Z_1 Z_2  Z_3 \right)^{-{2 \over 3}} (dt+k)^2
+ \left( Z_1 Z_2 Z_3\right)^{1 \over 3} \, ds_4^2  \\+ \left(Z_2
Z_3 Z_1^{-2}  \right)^{1 \over 3} (dx_5^2+dx_6^2) + \left( Z_1 Z_3
Z_2^{-2} \right)^{1 \over 3} (dx_7^2+dx_8^2)  + \left(Z_1 Z_2
Z_3^{-2} \right)^{1 \over 3} (dx_9^2+dx_{10}^2) \,,
\label{11Dmetric}
\end{multline}
where $ds_4^2$ is a four-dimensional hyper-K\"{a}hler metric
\cite{Gauntlett:2002nw,Gutowski:2004yv,Gauntlett:2004qy,Bena:2004de} \footnote{This
  metric can have regions of signature $+4$ and signature $-4$
  \cite{Giusto:2004kj,Bena:2005va,Berglund:2005vb,Saxena:2005uk}, and for this
  reason we usually refer to it as ambipolar.}. If one imposes in
addition an extra triholomorphic $U(1)$ isometry, this space has to be
a Gibbons-Hawking (GH) space  \cite{Gibbons:1987sp}:

\be
ds_4^2 = V^{-1}( d\psi + A )^2 + V ds_3^2
\label{metricGH}
\ee
with $ds_3^2$ the flat three-dimensional metric and $\vec{\nabla}V = \vec{\nabla} \times \vec{A}$.

The solution has a
non-trivial three-form potential, sourced both by the M2 branes
(electrically) and by the M5 dipole branes (magnetically):
\begin{equation}
\mathcal{A} = A^{(1)}\wedge dx_5 \wedge dx_6 + A^{(2)}\wedge dx_7
\wedge dx_8 + A^{(3)}\wedge dx_9 \wedge dx_{10}
\label{11Dthreeform}.
\end{equation}
The magnetic contributions can be separated from the electric ones
by defining the ``magnetic field strengths:"
\begin{equation}
\Theta^{(I)} \equiv dA^{(I)} ~+~
d\left( \frac{(dt+k)}{Z_I}\right)\,, \qquad I=1,2,3.
\end{equation}
Finding supersymmetric supergravity solutions for bubbles geometries  reduces to solving
the following system of BPS equations on the GH base \cite{Bena:2004de}:
\begin{equation}
\begin{array}{l}
\Theta^{(I)} = \star_4 \Theta^{(I)}\,, \\\\
\Box Z_{I} = \frac{1}{2}C_{IJK} \star_{4} (
\Theta^{(J)}\wedge
\Theta^{(K)})\,, \\\\
dk + \star_4 dk = Z_{I} \Theta^{I} \,.
\end{array} \label{BPSequations}
\end{equation}
In these equations, $\star_4$ is the Hodge dual in the four-dimensional base space and
$C_{IJK}=|\epsilon_{IJK}|$.  The operator, $\Box $, is the Laplacian in the four-dimensional metric.

Our purpose is to solve this in a similar manner to that described in
\cite{Bena:2004de}, except that we want to allow varying electric charge densities.
We will also later focus on three-charge geometries containing supertubes and so we will work in the D1-D5-P duality frame (where supertubes source smooth solutions).
In this frame, the metric takes the form:
\begin{eqnarray}
ds^2_{IIB}  = - \frac{1}{Z_3\sqrt{Z_1Z_2}}\,(dt+k)^2  &+&
\sqrt{Z_1Z_2}\, ds^2_4 +  \frac{Z_3}{\sqrt{Z_1Z_2}}\,(dz+A^{(3)})^2  \\
&+&   \sqrt{\frac{Z_1}{Z_2}}\, (dx_5^2+dx_6^2+dx_7^2+dx_8^2) \,,
\label{D1D5Pmetric}
\end{eqnarray}
where  $ds^2_4$ is still the metric of the GH base.  In the M-theory frame, the electromagnetic fields all appear in the  $3$-form Maxwell potential (\ref{11Dthreeform}) while in the IIB D1-D5-P frame one of these fields has become part of the metric (\ref{D1D5Pmetric}).

The first step is to solve the first BPS equation in exactly the same manner as for GH base spaces, \cite{Bena:2005ni,Bena:2005va, Berglund:2005vb,Bena:2007kg}  and for this it is convenient to introduce the vielbeins:
\begin{equation}
\hat e^1~=~ V^{-{1\over 2}}\, (d\psi ~+~ A) \,, \qquad \hat
e^{a+1} ~=~ V^{1\over 2}\, dy^a \,, \quad a=1,2,3 \,,
\label{GHvierbeins}
\end{equation}
and the two-forms:
\begin{equation}
\Omega_\pm^{(a)} ~\equiv~ \hat e^1  \wedge \hat
e^{a+1} ~\pm~ \coeff{1}{2}\, \epsilon_{abc}\,\hat e^{b+1}  \wedge
\hat e^{c+1} \,, \qquad a =1,2,3\,.\
\label{twoforms}
\end{equation}
One then takes
\begin{equation}
\Theta^{(I)} ~=~ - \sum_{a=1}^3 \, \left(\partial_a \left(
V^{-1}\, K^I \right)\right) \, \Omega_+^{(a)}  \,, \label{GHdipoleforms}
\end{equation}
for some harmonic functions, $K^I$, on the $\IR^3$ base of the GH space.
We specifically require the magnetic fluxes to be independent of $\psi$.  These field strengths have potentials:
\begin{equation}
 B^{(I)}=V^{-1} K^{I} \, (d \psi + A) ~+~ \vec{\xi}^{(I)}\cdot d\vec{y} \,,
 \qquad \vec \nabla \times \vec \xi^{(I)}  ~\equiv~ - \vec \nabla K^I \,,
\label{vecpotdefns}
\end{equation}
where $ \Theta^{(I)} = d B^{(I)}$.  

The source of the second BPS equation is independent of $\psi$ and so the inhomogeneous solutions for the functions $Z_I$  follow the standard form:
\begin{equation}
Z_I  ~=~ \coeff{1}{2}\, C_{IJK}  V^{-1} K^{J}K^{K}   ~+~ L_I  \,,
\label{ZIform}
\end{equation}
and the functions $L_I$ are still required to be harmonic but are now going to be allowed to depend upon all variables including the fiber.
Thus we have 
\begin{equation}
\Box L_I  ~=~0  \,.
\label{Lharmcond}
\end{equation}

As before, there is a natural Ansatz for the angular momentum one-form $k$:  
\begin{equation}
k ~=~ \mu (d\psi+ A) + \vec{\omega}\cdot d\vec{y} \,.
\label{kansatz}
\end{equation}
The third BPS equation then reduces to:
\begin{eqnarray}
( \mu \vec \nabla V - V \vec \nabla \mu  ) ~+~  V\big[\partial_\psi \vec \omega ~+~   \vec A \, \partial_\psi \mu \big]  &+&   \vec \nabla \times \vec \omega ~+~    (\partial_\psi  \vec \omega \times \vec A) \nonumber \\
  &=& -  V\, \sum_{I=1}^3 \, Z_I \, \vec \nabla \big(V^{-1} K^I  \big) \,,
\label{keqn}
\end{eqnarray}
where $\vec A$ is the vector field in the fibration of the GH metric and satisfies $\vec \nabla \times \vec A =\vec \nabla V$.

If one defines the covariant derivative:
\begin{equation}
\vec {\cal D} ~\equiv~ \vec \nabla ~-~   \vec A \,\partial_\psi   \,,
\label{covD}
\end{equation}
then one can write (\ref{keqn}) as
\begin{equation}
( \mu \vec {\cal D} V - V\vec {\cal D}  \mu  ) ~+~      \vec {\cal D}  \times \vec \omega ~+~    V \partial_\psi  \vec \omega  ~=~ -  V\, \sum_{I=1}^3 \, Z_I \, \vec \nabla \big(V^{-1} K^I  \big) \,.
\label{covkeqn}
\end{equation}
Moreover, the four-dimensional Laplacian can be written:
\begin{equation}
\Box  F ~=~ V^{-1} \big[ V^2 \, \partial_\psi^2  F   ~+~ \vec {\cal D}  \cdot  \vec {\cal D}    F \big] \,.
\label{Lapl}
\end{equation}

The third BPS equation has a gauge invariance:  $k \to k + df$ and this reduces to:
\begin{equation}
\mu \to \mu ~+~ \partial_\psi f \,, \qquad  \vec \omega \to  \vec \omega ~+~ \vec {\cal D} f \,,
\label{fgaugetrf}
\end{equation}
The Lorentz gauge-fixing condition, $d\star_4k =0$, reduces to
\begin{equation}
V^2 \, \partial_\psi \mu   ~+~\vec {\cal D}  \cdot \vec \omega  ~=~ 0 \,,
\label{Lorgauge}
\end{equation}
and we will adopt this gauge throughout.

Now take the covariant divergence, using $\vec {\cal D}$,  of (\ref{covkeqn}) and use the Lorentz gauge choice, and one obtains:
\begin{equation}
V^{^2}\, \Box  \mu  ~=~   \vec {\cal D} \cdot \Big( V\, \sum_{I=1}^3 \, Z_I \, \vec {\cal D} \big(V^{-1} K^I  \big) \Big) \,.
\label{mueqn}
\end{equation}
Remarkably enough, this equation is still solved by:
\begin{equation}
\mu ~=~ \coeff{1}{6}\,  V^{-2}C_{IJK}  K^{I}K^{J}K^{K} ~+~ \coeff{1}{2}\, V^{-1} K^{I}L_{I}  ~+~  M\,,
\label{muform}
\end{equation}
where, once again, $M$ is a harmonic function in four dimensions.
Finally, we can use this solution back in (\ref{covkeqn}) to simplify the right-hand side to obtain:
\begin{equation}
\vec {\cal D} \times \vec \omega ~+~ V \partial_\psi \vec \omega ~=~ V \vec {\cal D} M - M\vec {\cal D} V
+\frac{1}{2} \big( K^{I} \vec {\cal D} L_{I} - L_{I}   \vec {\cal D} K^{I} \big).
\label{omegaeqn}
\end{equation}
Once again one sees the emergence of the familiar symplectic form on the right-hand side of this equation.  One can also verify that the covariant divergence (using $\vec {\cal D}$) generates an identity that is trivially satisfied as a consequence of  $ \vec\nabla V =  \vec \nabla \times \vec A$, (\ref{Lorgauge}),  (\ref{muform})  and
\begin{equation}
\Box L_I ~=~\Box   M ~=~ 0 \,.
\label{harmonicLM}
\end{equation}
While it might be possible to find, we do not have an explicit closed form for $\vec \omega$, but we will not need it.

We have thus generalized the results of \cite{Gauntlett:2004qy,Bena:2005ni,Bena:2005va,Berglund:2005vb} to completely general, fluctuating electric charge densities.

%%%%%%%%%%%%%%%%%%%%%%%%%%%%%%%%%%%%%
\section{Scalar Green functions on a GH space}
%%%%%%%%%%%%%%%%%%%%%%%%%%%%%%%%%%%%%
\label{Greensfunctions}

In order to solve \eqref{harmonicLM} for varying charge densities, one needs the scalar Green function on a GH space. When the GH space is a regular Euclidean space (the function $V$ is everywhere positive), the Green function is well-known \cite{Page:1979ga}.  However, as we explained in the Introduction, when the GH space is ambipolar finding the Green function is not so straightforward, and the simple analytic continuation of the Green function on a regular GH space does not produce a physical ambipolar Green function. We therefore turn to an indirect method for constructing this Green function: we first construct the Green function for a five-dimensional solution that has an ambipolar base space: $AdS_3 \times S^2$, and then we reduce this Green function.

This section reviews some of the known material about Green functions on GH spaces and the relationship between $AdS_3 \times S^3$ and an ambipolar two-centered GH space.  Much of the rest of this section contains the (highly technical) derivation of the Green function that we will subsequently use. Those who wish to skip most the technical details can get a good idea of the general strategy from Section \ref{Gstrategy}.  The key physical results that we will need about the Green function are contained in Section 
\ref{Gasymptotics}, which discusses the behavior of the Green function in various limits.  In particular, Section \ref{GatGHpts} contains the essential result that demonstrates that  the bubble equations at the GH fibers (but not at the supertube itself) are unchanged by the fluctuations and only depend upon the total charge of the fluctuating component.

%%%%%%%%%%%%%%%%%%%%%%%%%%%%%%%%%%%%%
\subsection{Review of known Green functions on regular GH spaces}
%%%%%%%%%%%%%%%%%%%%%%%%%%%%%%%%%%%%%%

The Green function on a positive definite, multi-center Taub-NUT space have been known for some time.  Consider the metric (\ref{metricGH}) with $V$ harmonic and everywhere positive:
\be
V=\sum_{n}{q_n\over |\vec x - \vec x_n|} \,,
\ee
with $q_n \ge 0$.
The Green function was found by Page in \cite{Page:1979ga}, and is given by:
\be
G(\psi,\vec x;\psi',\vec x')={1\over 16\pi^2 \Delta} {\sinh U\over \cosh U -\cos T}\,,
\label{PageGF}
\ee
where
\be
\Delta \equiv |\vec{x}-\vec{x}'|
\ee
and
\bea
T&\equiv&{\psi-\psi'\over 2}+\sum_n q_n \arctan \Bigl[{\cos{\theta_n
+\theta'_n\over 2}\over \cos{\theta_n-\theta'_n \over 2}}
\, \tan {\phi_n-\phi'_n \over 2}\Bigr]\nonumber\\
U&\equiv&{1\over 2}\sum_n q_n \log
\Bigl[{r_n+r'_n+\Delta\over r_n+r'_n-\Delta}\Bigr]~, \label{UTeq}
\eea
and for brevity we have introduced spherical polar coordinates $(r_n,\theta_n,\phi_n)$ around each pole at position $\vec x_n$.
The only singularity of this Green function is at $U=T=0$, which for regular GH spaces with $q_n \ge 0$ only happens when the points are coincident: $\psi=\psi'$ and $\vec x=\vec x'$.

By integrating this Green function against the fiber $\psi$, one obtains the trivial Green function on $\IR^3$:
\be
\int_0^{4\pi} \!d\psi\, G = {1\over 4\pi \Delta}\,.
\ee

One might hope that that equation (\ref{PageGF}) would also give the Green function for ambipolar spaces where some of the $q_n$ are
negative, and indeed the Green function one obtains (which we refer to as the {\it Page Green function}) satisfies the appropriate differential equation. As before, this Green function is singular when $U=T=0$, but in an ambipolar space this does not only
happen when the two points are coincident.  Indeed, as one can see from Eq (\ref{UTeq}), for a fixed  $(\psi',  \vec x')$,  the points $(\psi,  \vec x)$ defined by  $U=T=0$ belong to a codimension-two hypersurface, and in general $U$ and $T$ can be both positive and negative.

One could also calculate the integral of the ambipolar Page Green function over  the fiber $\psi$:
\be
\int_0^{4\pi} \!d\psi \,G_{Page} = {U \over |U|}{1\over 4\pi \Delta}
\ee
 and see that this integral is discontinuous along the surface $U=0$.

One can also try to construct a physical  Green function on an ambipolar space by piecing  the Page Green functions on the patches $U>0$ and $U<0$:
\be
G_{patch} = {1\over 16\pi^2 \Delta} {\sinh|U| \over \cosh U -\cos T}\,.
\ee
The $\psi$ integral of $G_{patch}$ then gives the correct three-dimensional propagator, but this function has a non-physical cusp across the $U=0$ hypersurface. Apart from $G_{Page}$ and $G_{patch}$, there does not seem to be any natural guess for a physical continuation of the multi-center Taub-NUT Green function to ambi-polar spaces, and hence we do not have a direct method to construct this Green function.

Having failed in the brute-force construction, we now take the less direct road by remembering the ultimate goal of the overall construction:  A smooth five-dimensional geometry.   We therefore construct the Green function for a particular family of ambipolar spaces by dimensionally reducing the Green function of the smooth five-dimensional solutions built using these spaces as a base.

%%%%%%%%%%%%%%%%%%%%%%%%%%%%%%%%%%%%%
\subsection{Preliminaries: the simplest metric with an ambi-polar base, global $AdS_3 \times S^2$}
%%%%%%%%%%%%%%%%%%%%%%%%%%%%%%%%%%%%%

It has been noted by several authors (see, for example, \cite{Denef:2007yt,deBoer:2008fk}) that the five-dimensional metric metric arising from an ambi-polar Eguchi-Hanson metric with GH charges $+q$ and $-q$ is, in fact, a $\ZZ_q$ quotient of global $AdS_3 \times S^2$. For simplicity, we will locate these GH charges on the $z$-axis at $ z= \pm a$ and define:
\begin{equation}
r_\pm ~\equiv~   \sqrt{\rho^2 ~+~ (z\mp a)^2 } \,,
\end{equation}
where $(z, \rho, \phi)$  are cylindrical polar coordinates on the $\IR^3$ base.
The harmonic functions for this solution are
\begin{eqnarray}
V &=&  q\, \Big({ 1 \over r_+} ~-~ {1 \over r_-} \Big)\,, \qquad  K ~=~
k\, \Big({ 1 \over r_+} ~+~ {1 \over r_-} \Big)  \,,  \\
\qquad  L&=&- {k^2 \over q} \Big({ 1 \over r_+} ~-~ {1 \over r_-} \Big) \,, \qquad  M ~=~
- {2\, k^3 \over a\, q^2}~+~ \frac{1}{2} \,{k^3 \over q^2} \Big({ 1 \over r_+} ~+~
{1 \over r_-} \Big)\,,
\end{eqnarray}
where the constant in $M$ has been chosen so as to make the metric regular at infinity.

The vector potentials for this solution are then:
\begin{equation}
A ~=~   q\, \Big({(z -a) \over r_+} - {(z +a) \over r_-} \Big) \, d\phi  \,, \qquad \omega ~=~ -{2\, k^3 \over a\, q} \,
{\rho^2 + (z-a +r_+)(z+a - r_-)  \over r_+ \, r_-}  \, d\phi  \,.
\end{equation}
The five-dimensional metric is then:
\begin{equation}
ds_5^2 ~\equiv~ - Z^{-2} \big(dt+ \mu (d\psi+A) + \omega\big)^2 ~+~
Z\, \big(   V^{-1} (d\psi+A)^2 ~+~ V(d\rho^2 + \rho^2 d\phi^2 + dz^2)  \big)\,,
\label{fivemetric}
\end{equation}
where
\begin{eqnarray}
Z &=&  V^{-1} K^2 + L ~=~ - {4\, k^2 \over q}  \, { 1 \over (r_+ - r_- )}\,,  \\
\mu &=&  V^{-2} K^3 +  \coeff{3}{2}\, V^{-1} K \, L + M ~=~ {4\, k^3 \over q^2}  \,
{  (r_+ + r_- ) \over (r_+ - r_- )^2} ~-~  {2\, k^3 \over a\, q^2} \,. \nonumber
\end{eqnarray}

To map this onto a more standard form of $AdS_3 \times S^2$ one must make a
transformation to oblate spheroidal coordinates like the one employed in \cite{Prasad:1979kg}
to map positive-definite two-centered GH space onto the Eguchi-Hanson form:
\begin{equation}
 z =  a\, \cosh 2\xi \,\cos \theta \,, \qquad  \rho =  a\, \sinh 2\xi \, \sin \theta \,, \qquad
 \xi \ge 0\,, \ \ 0 \le \theta \le \pi \,.
  \label{coordsa}
 \end{equation}
In particular, one has $r_\pm =  a (\cosh 2\xi  \mp \cos \theta)$. (Note that we have taken the argument of the hyperbolic functions to be $2 \xi$ for later convenience.)  One then rescales and shifts the remaining variables according to:
\begin{equation}
 \tau ~\equiv~   \coeff{a\, q}{8\, k^3}\,  t \,, \qquad \varphi_1 ~\equiv~   \coeff{1}{2\,q} \, \psi -
 \coeff{a\, q}{8\, k^3}\,  t   \,, \qquad \varphi_2 ~\equiv~ \phi -   \coeff{1}{2\,q} \, \psi +
 \coeff{a\, q}{4\, k^3}\,  t   \,,
 \label{coordsb}
 \end{equation}
and the five-dimensional metric takes the standard $AdS_3 \times S^2$ form:
\begin{equation}
ds_5^2 ~\equiv~ R_1^2 \big[ - \cosh^2\xi \,  d\tau^2 + d\xi^2 +  \sinh^2 \xi \, d\varphi_1^2 \big] ~+~  R_2^2 \big[   d \theta ^2 + \sin^2\theta  \, d\varphi_2^2 \big]  \,,
 \label{AdS3S2}
 \end{equation}
with
\begin{equation}
  R_1~=~  2 R_2 ~=~ 4 k \,.
 \label{Radii}
 \end{equation}
Note that the metric involves {\it global} $AdS_3$ with $-\infty < \tau < \infty$.

One should also recall that the GH fiber coordinate has period $4 \pi$
and therefore, for $|q|=1$, the angles, $\varphi_j$, both have periods
$2 \pi$.  For $q \ne 1$, the $\ZZ_{|q|}$ orbifold associated with the
GH points emerges as a simultaneous $\ZZ_{|q|}$ quotient on the
longitudes of the $AdS_3$ and $S^2$.
 
It is instructive to observe that a particle that is at a fixed spatial point in the GH base has a world-line with constant $\xi$ and $\theta$ and
\begin{equation}
 {d  \varphi_1 \over d \tau} ~=~   -1 \,, \qquad {d  \varphi_2 \over d \tau} ~=~ - 2   \qquad
 \Rightarrow \qquad ds_5^2 ~=~ - 16 k^2 \cos^2 \theta \, d\tau^2\,.
 \label{staticworldline}
 \end{equation}
This is the world-line of a observer rotating around the $AdS_3$ and
 $S^2$ simultaneously along a curve that is null for $\theta ={\pi
   \over 2}$ and otherwise time-like.  The surface $\theta ={\pi \over
   2}$ is the simply the critical surface defined by $V=0$ in the GH
 base.  As has been noted elsewhere, and is evident from
 (\ref{AdS3S2}), the five-dimensional metric is completely regular
 across this surface and the only artifact of an apparently singular
 four-dimensional base is that a stationary particle on this base is
 simply following a null curve on the critical surface in the
 five-dimensional geometry.

%%%%%%%%%%%%%%%%%%%
\subsection{The strategy: reducing a Green function from five to four dimensions}
\label{Gstrategy}
%%%%%%%%%%%%%%%%%%%

Our goal here is to obtain the scalar Green function for the
ambi-polar GH base considered above, and construct this Green function
in such a manner that it can be used to create five-dimensional solutions by the methods outlined in
\cite{Bena:2004de}.  One can obtain such a
Green function from the Green function for $AdS_3 \times S^2$ by
integrating it along the time coordinate, $t$, in (\ref{fivemetric}).

To be specific, suppose that we have a function, ${\cal G}(x;x')$, that satisfies
\begin{equation}
\Box_x \, {\cal G}(x;x') ~=~ \frac{1}{\sqrt{-g}}\,\delta(x;x')  \,,
 \label{fivedimG}
 \end{equation}
on $AdS_3 \times S^2$.  Let $x = (y,t)$, where $y$ stands for the coordinates on the GH base.  Write the metric (\ref{fivemetric}) as:
\begin{equation}
ds_5^2 ~\equiv~ - Z^{-2} \big(dt+ k_m dy^m \big)^2 ~+~  Z\,  h_{m n}\,dy^m dy^n \,,
\label{genfivemetric}
\end{equation}
and observe that  (\ref{fivedimG}) is equivalent to:
\begin{equation}
\ -Z^{-3} \partial_t^2 {\cal G}(x;x') ~+~  \frac{1}{ \sqrt{h}}  (\partial_m  - k_m \partial_t) \,
\big( \sqrt{h} \,h^{mn}  (\partial_n  - k_n \partial_t) \,{\cal G}(x;x')  \big) ~=~ \frac{1}{\sqrt{h}}\, \delta(x;x')    \,.
 \label{Boxsimp}
 \end{equation}

Since all the metric coefficients are time-independent, one can take all the time derivatives to be total derivatives of the various terms in (\ref{Boxsimp}).   Integrating this identity over $t$ then yields the identity of the form:
\begin{equation}
\Box_y \, G (y;y')  ~=~ \frac{1}{ \sqrt{h}}  \partial_m \,
\big( \sqrt{h} \,h^{mn}   \partial_n   \, G (y;y')  \big) ~=~ \frac{1}{\sqrt{h}}\, \delta(y; y')    \,,
 \label{RedBox}
 \end{equation}
provided that one can drop the boundary terms from the $t$  integration.   We will construct the Green  function with a periodic time variable and then analytically continue using $\tau - \tau' = i \lambda$.  With this prescription the boundary terms will indeed vanish.  We therefore define
\begin{equation}
G (y;y')  ~\equiv~  \int_{-\infty}^{\infty}\,  d \lambda\ {\cal G}(y, \tau = \tau'+ i \lambda ;y', \tau')  \,,
 \label{Gdefn}
 \end{equation}
and this will give us the required Green function on the GH base.  One should note that because of the off-diagonal ``angular momentum'' terms in the metric (\ref{genfivemetric}), the analytic continuation of ${\cal G}(x;x')$ will not produce a real integrand in (\ref{Gdefn}), however, the imaginary parts will disappear as boundary terms in the integral.  Periodic identification of the time coordinate, $\tau$, introduces an infinite number of image sources into the original $AdS_3$ space and one might be concerned that this will generate spurious sources in $G (y;y')$.  However this is not a problem because we are integrating over the periodic variable and the periodicity is commensurate with that of the $\varphi_j$.

%%%%%%%%%%%%%%%%%%%
\subsection{The scalar Green function on $AdS_3 \times S^2$ }
%%%%%%%%%%%%%%%%%%%

The scalar Green functions for $AdS_3 \times S^2$ has been extensively
discussed in \cite{Dorn:2003au} and here we simply follow their
prescription.  It also turns out that the particular $AdS_3 \times
S^2$ given by (\ref{Radii}) has some extremely nice properties with
respect to the procedures of \cite{Dorn:2003au}.

Consider the more general equation:
\begin{equation}
(\Box_{AdS_3} +  \Box_{S^2} - M^2) \, {\cal G}(x;x') ~=~ \frac{i}{\sqrt{-g}}\,\delta(x;x')  \,,
 \label{Gfneqn}
 \end{equation}
 where the factor of $i$ has been included so that, upon passing to
 the Euclidean version, the Green function is real.  Expand the Green
 functions into spherical harmonics on $S^2$ one obtains:
\begin{equation}
  {\cal G}(x;x') ~=~  \frac{1}{R_2^2} \,\sum_{\ell,m} {\cal G}_{\ell,m} (w;w')  \, Y_{\ell,m} (\theta,\varphi_2) Y^*_{\ell,m} (\theta',\varphi'_2)\,,
 \label{Gfnexpansion}
 \end{equation}
where $w$ and $w'$ are coordinates on $AdS_3$ and
\begin{equation}
\bigg(\Box_{AdS_3}  - M^2 -  \frac{\ell(\ell+1)}{R_2^2} \bigg) \, {\cal G}_{\ell, m}(w;w') ~=~ \frac{i}{\sqrt{-g_{AdS}}}\,\delta(w;w')  \,.
 \label{Glmfneqn}
 \end{equation}

Since the functions, ${\cal G}_{\ell,m}$, do not, in fact, depend upon $m$, we may perform the sum over $m$ to obtain:
\begin{equation}
{\cal G}(x;x') ~=~  \frac{1}{4\pi R_2^2} \,\sum_{\ell=0}^\infty {\cal G}_{\ell} (w;w')  \, (2 \ell +1)  P_\ell(\cos \gamma)  \,,
 \label{Gfnexpsimp}
 \end{equation}
where $\gamma$ is the great-circle angular separation on the $S^2$:
\begin{equation}
\cos \gamma ~\equiv~  \cos \theta \, \cos \theta' +  \sin \theta \, \sin \theta'  \cos(\varphi_2 -\varphi_2') \,.
 \label{gammadefn}
 \end{equation}

One can now solve the equation (\ref{Glmfneqn}) for  ${\cal G}_{\ell}$ in terms of the ``chordal distance'' on $AdS_3$.  Consider the hyperboloid:
\begin{equation}
(X_0)^2 +  (X_3)^2  -  (X_1)^2 -  (X_2)^2  ~=~ R_1^2\,,
 \label{AdShyp}
 \end{equation}
with the metric induced from
\begin{equation}
ds^2 ~=~ - dX_0^2 - dX_3^2  + dX_1^2  + dX_2^2  \,.
 \label{flatmet}
 \end{equation}
With the coordinate change:
\begin{eqnarray}
X_0 &=& R_1 \cosh \xi \, \cos \tau \,, \quad X_3 ~=~  R_1 \cosh \xi \, \sin \tau \,, \\
 X_1 &=&   R_1 \sinh \xi \, \cos \varphi_1 \,, \quad X_2 ~=~ R_1\sinh \xi \, \sin \varphi_1 \,,
 \label{AdShypcoords}
 \end{eqnarray}
the induced metric becomes precisely the $AdS_3$ factor of (\ref{AdS3S2}) {\it except} that here the time coordinate, $\tau$, is periodically identified with period $2 \pi$.   As we noted above, this periodic identification is consistent with the the world-lines (\ref{staticworldline}) over which we wish to integrate.

The chordal distance is given by:
\begin{equation}
v ~\equiv~  - (X_0 -X_0' )^2 - (X_3 -X_3' )^2 + (X_1 -X_1' )^2+ (X_2 -X_2' )^2 ~=~
2R_1^2 \, ( \zeta  ~-~ 1) \,,
 \label{AdSchord}
 \end{equation}
where
\begin{equation}
\zeta  ~\equiv~   \cosh \xi  \cosh \xi'  \cos (\tau - \tau')  ~-~  \sinh \xi  \sinh \xi'  \cos (\varphi_1 -\varphi_1')\,.
 \label{zetadefn}
 \end{equation}
One can then show that  \cite{Dorn:2003au}:
\begin{equation}
{\cal G}_\ell ~=~    \frac{1}{2^{\Delta +1} \pi R_1}  \, \zeta^{- \Delta} F \big(\coeff{1}{2} \Delta, \coeff{1}{2} \Delta  + \coeff{1}{2} ;\Delta; \zeta^{-2} \big)  ~=~
 \frac{1}{4\pi R_1} \, {\big (\zeta + \sqrt{\zeta^2 -1}\big )^{1- \Delta} \over  \sqrt{\zeta^2 -1} }  \,,
 \label{Gellres}
 \end{equation}
where $F$ is a hypergeometric function and
\begin{equation}
\Delta~\equiv~   1+ \frac{R_1}{R_2}  \,\sqrt{\frac{R_2^2}{R_1^2}  + \ell(\ell+1) + M^2 R_2^2}\,.
 \label{Deltawtdefn}
 \end{equation}

We want the Green function for $M=0$ and $R_1 = 2 R_2$ for which there is a remarkable simplification: $\Delta = 2 (\ell +1)$ and the sum in (\ref{Gfnexpsimp}) can be performed explicitly.  Let  $\eta \equiv  \big(\zeta + \sqrt{\zeta^2 -1} \big)^{-1}=\zeta - \sqrt{\zeta^2 -1} $ and observe that  (\ref{Gfnexpsimp}) has the form:
\begin{eqnarray}
{\cal G} &=&    \frac{1}{8 \pi^2 R_1  R_2^2} \,{\eta \over 1 -\eta^2} \, \sum_{\ell =0}^\infty (2 \ell +1) \, \eta^{2 \ell +1} \, P_\ell (\cos \gamma)   \nonumber \\
&=&    \frac{1}{8 \pi^2 R_1  R_2^2}  \,{\eta^2  \over 1 -\eta^2} \,    {d \over d \eta}
\sum_{\ell =0}^\infty \eta^{2 \ell +1} \, P_\ell (\cos \gamma) \nonumber \\
 &=&    \frac{1}{8 \pi^2 R_1  R_2^2}\,{\eta^2  \over 1 -\eta^2} \,   {d \over d \eta} \big(\eta \, \big(  1 + \eta^4 - 2 \eta^2 \cos \gamma \big)^{-{1 \over 2}}  \big) \nonumber \\
&=&  \frac{1}{8 \pi^2 R_1 R_2^2} \,  {( \eta  +  \eta^{-1})\over ( \eta^2+ \eta^{-2} - 2  \cos \gamma )^{3\over 2} }\,.
\label{sumseries}
 \end{eqnarray}

We therefore find
\begin{equation}
{\cal G} ~=~    \frac{1}{8 \sqrt{2}\, \pi^2 R_1  R_2^2} \,  { \zeta \over (2 \zeta ^2 - (1+  \cos \gamma) )^{3\over 2} }\,,
 \label{Gfnans}
 \end{equation}
where $\gamma$ and $\zeta$ are given by (\ref{gammadefn})  and (\ref{zetadefn}).  The result is thus a relatively simple combination of chordal distances on  $AdS_3$ and $S^2$.

One should note that in the Euclideanized $AdS_3$ with $\tau \to i \tau$, it follows from (\ref{zetadefn}) that $\zeta \ge 1$ with equality if and only if $\xi = \xi'$, $\tau= \tau'$ and $\varphi_1 = \varphi_1'$.  The denominator of  (\ref{Gfnans})  only vanishes if $\zeta =1$ and $\gamma =0$, which means that only singularity of ${\cal G}$ occurs when the points coincide on the Euclideanized background.  In addition, $\zeta > 1$ means that $\eta <1$ and this is the condition for convergence of the series in (\ref{sumseries}).

We have therefore constructed the scalar Green function on Lorentzian $AdS_3 \times S^2$ with periodic time and seen that it has the appropriate singular structure when Wick rotated to Euclidean $AdS_3 \times S^2$.
To get to the Green function on the  GH base we have to use the transformation (\ref{coordsb}) and this will lead to a somewhat different analytic continuation.  We now investigate this in more detail.

%%%%%%%%%%%%%%%%%%%
\subsection{Reducing the scalar Green function to the GH base}
%%%%%%%%%%%%%%%%%%%

To get the requisite Green function, $G (y;y')  $, on the GH base, we have to make the changes of variable:
\begin{equation}
\varphi_1 ~=~   \coeff{1}{2\,q} \, \psi -  \tau   \,, \qquad \varphi_2 ~\equiv~ \phi -   \coeff{1}{2\,q} \, \psi + 2 \tau  \,,
 \label{coordsc}
 \end{equation}
make the analytic continuation $\tau - \tau' = i \lambda$ and perform the integral (\ref{Gdefn}).  Simplifying the normalization for the present, we have:
\begin{eqnarray}
{\cal I} &\equiv&  8 \sqrt{2}\, \pi^2 R_1  R_2^2  \,  \int_{-\infty}^{\infty}    {\cal G} \,d \lambda ~=~ \int_{-\infty}^{\infty}     { \zeta \over (2 \zeta ^2 - (1+  \cos \gamma) )^{3\over 2} } \,d \lambda   \,,  \\
 &=&  \int_{-\infty}^{\infty}   {a_1 \cosh \lambda  + i b_1  \sinh \lambda \over \big( a_2 \cosh 2\lambda  + i b_2  \sinh 2\lambda + a_0\big)^{3 \over 2}  }  \,d \lambda\,,
 \label{int1}
 \end{eqnarray}
where the $a_i$ and $b_j$ are {\it real} variables.  It turns out that this integral is elementary and it is given by:
\begin{equation}
{\cal I} ~=~  {p \over \sqrt{\coeff{1}{2}(a_2 +  i b_2)}}   +  {\bar p \over \sqrt{\coeff{1}{2}(a_2 -  i b_2)}}  \,,
 \label{intans1}
 \end{equation}
where
\begin{equation}
p ~\equiv~ \coeff{1}{2} \,(a_0^2 - (a_2^2 + b_2^2))^{-1}\big[ (a_0 -a_2) a_1 - b_1 b_2 + i ( (a_0 + a_2) b_1 - a_1 b_2 ) \big] \,,
 \label{pdefn}
 \end{equation}
and $\bar p$ is its complex conjugate.  The result, while apparently rather simple, is an extremely compressed form of the answer we are seeking and we first rewrite it in terms of the coordinates of the points. 

Define the following combinations of coordinates:
\begin{eqnarray}
u  &\equiv&  \cosh \xi  \cosh \xi' \,,  \quad v ~\equiv~ \sinh \xi  \sinh  \xi'  \,, \quad
x ~\equiv~    \cos \theta  \cos \theta' \,, \nonumber  \\ 
y  &\equiv&    \sin \theta  \sin \theta'   \,, \quad
 z  ~\equiv~ e^{{i \over 2 q} (\psi - \psi')}   \,, \qquad  \chi  ~\equiv~   (\phi - \phi') \,,
 \label{params1}
 \end{eqnarray}
then one has
\begin{eqnarray}
a_1 &=&   u - \coeff{1}{2}  v\, ( z+ z^{-1})\,, \quad b_1 ~=~   \coeff{i}{2}  v\,  ( z- z^{-1})\,,  \nonumber \\
a_0   &=&  a_1^2 +  b_1^2   -  (1+x)  \,,  \quad a_2 + i b_2 ~=~  (u - v z)^2  - y e^{-i \chi} z  \,.
 \label{params2}
 \end{eqnarray}
It is also convenient to take:
\begin{eqnarray}
\Lambda  &\equiv&  z\,(a_0^2 - (a_2^2 + b_2^2)) \,,  \quad
m ~\equiv~   2(1+x) u v +  y  (u^2   e^{-i \chi}  + v^2  e^{ i \chi})\,,  \label{functions1} \\
\nu  &\equiv&  (a_0 -a_2) a_1 - b_1 b_2 + i ( (a_0 + a_2) b_1 - a_1 b_2 ) \,, \label{functions2}\\
{\cal P}  &\equiv&   \big(2(u-v)^2 - (1+x +y) \big) \big(2(u+v)^2 - (1+x - y) \big)+ 8 u v y (1 -\cos \chi)\,,   \label{functions3}\\
z_\pm  &\equiv&  \coeff{1}{2 m} \big[y^2 - (1+x)^2  + 2(u^2 + v^2)(1+x) + 4u v y \cos \chi \pm \sqrt{ (1+x)^2-y^2 }\, \sqrt{{\cal P}} \, \big] \,.
 \label{functions4}
 \end{eqnarray}
Adjusting the overall normalization, the desired Green function is then:
\begin{equation}
\widehat G ~\equiv~  8  \, \pi^2 {R_1  R_2^2  \over a} \, \int_{-\infty}^{\infty}    {\cal G} \,d \lambda~=~  {\rm Re}\Big[ { z\, \nu \over a\,\Lambda \sqrt{a_2 + i b_2} } \Big]\,.
\label{Gnice}
 \end{equation}
This is the Green function on the GH base and our task is now to unpack the details and understand the result.  As with all Green functions, the physics lies in the singular structure and asymptotic behaviors.  

%%%%%%%%%%%%%%%%%%%
\subsection{Properties of the Green function}
\label{Gprops}
%%%%%%%%%%%%%%%%%%%

While the Green function (\ref{Gnice}) is rather unintuitively expressed and represents a very complicated function of the physical variables, its overall structure can be understood fairly easily.  Here we will summarize the important features and relegate some more technical details to Appendix A.    Since $z$ is simply a phase, the physical domain has $|z| =1$, however because we are going to want to use residue calculus to evaluate integrals over physical values of $z$,  we are  going to want to understand  the behavior of this Green function in the domain $|z| \le 1$.    

Since all the constituent functions are analytic, the only potential sources of singularities are $\Lambda$ and  $\sqrt{a_2 + i b_2} $.   The bottom line will be that the function $\Lambda$ has essentially the same singularities as the Page Green function, while the remaining components play the crucial role of canceling the unphysical singularities or cusps at the critical ($V=0$) surface and at the dangerous ``image" surfaces, and giving a smooth physical Green function. 

 %%%%%%%%%%%%%%%%%%%
\subsubsection{Returning to the GH geometry}
%%%%%%%%%%%%%%%%%%%

We start by recasting our result, as far as we can, in terms of the original GH geometry.  Following \cite{Page:1979ga}, we define $r_\pm$, $r_\pm$, $\theta_\pm$ and $\theta_\pm'$ to be the radial distances and polar angles of the two points as measured from the GH points at $z = \pm a$, and we define the functions:
\begin{eqnarray}
T &\equiv&  \coeff{1}{2 q}(\psi - \psi') ~+~ \sum_{\pm} \pm \arctan\bigg [ {\cos \coeff{1}{2}(\theta_\pm +  \theta_\pm') \over \cos \coeff{1}{2}(\theta_\pm -   \theta_\pm') } \tan \coeff{1}{2} (\phi   -  \phi ') \bigg]    \\
U  &\equiv&  {1 \over 2} \,\sum_{\pm} \pm \log \bigg [     {r_\pm + r_\pm' + \Delta \over r_\pm + r_\pm' -  \Delta}  \bigg]    \,,
 \label{TUdefns}
 \end{eqnarray}
where
\begin{equation}
\Delta ~\equiv~ \big| \vec r ~-~ \vec r' \big|  ~=~ \big( r_\pm^2 + r_\pm'{}^2 - 2\,r_\pm r_\pm' \, ( \cos \theta_\pm \cos \theta_\pm' +  \sin \theta_\pm \sin \theta_\pm' \cos(\phi - \phi')  )\big)^{1 \over 2}\,.
 \label{Deltadefn}
 \end{equation}
Note that these functions are normalized differently from those of \cite{Page:1979ga} in that we have divided by the GH charge, $q$.

It is a rather tedious process to relate the foregoing, more geometric, functions and variables to those of the previous subsection.  The basic relationship is that $r_\pm =  a (\cosh 2\xi  \mp \cos \theta)$ and $r_\pm' =  a (\cosh 2\xi'  \mp \cos \theta')$.  Then by drawing triangles and finding expressions for various side lengths one can obtain:
\begin{eqnarray}
\sqrt{r_+  r_-} \,  \cos \coeff{1}{2}(\theta_+ -  \theta_-)    =  a \, \sinh 2 \xi  \,,  & \quad  \sqrt{r_+  r_-} \,  \sin \coeff{1}{2}(\theta_+ -  \theta_-) =   a \,  \sin \theta \,,   \\
\sqrt{r_+  r_-} \,  \cos \coeff{1}{2}(\theta_+ +  \theta_-)    =  a \, \sinh 2 \xi \cos \theta\,,  & \quad  \sqrt{r_+  r_-} \,  \sin \coeff{1}{2}(\theta_+ +  \theta_-) =   a \, \cosh 2 \xi \sin \theta     \,.
 \label{angleids}
 \end{eqnarray}

The function, $\Lambda$, is manifestly a quadratic in $z$, and one can easily show that it has the following form
\begin{equation}
\Lambda ~=~ m\, (z - z_+)(z -z_-)  \quad {\rm with} \quad z_+  z_- ~=~ {\bar m \over m}\,.
 \label{Lambdanice}
 \end{equation}
One can then establish the following identities:
\begin{eqnarray}
 \qquad e^{i T} &=& { m \over |m| } \, z  \,, \qquad
e^{\pm U} ~=~  { m \over |m| } \, z_\pm  \,,  \qquad \Delta   ~=~ a \, \sqrt{{\cal P}} \,,  \nonumber \\
|m| &=& {1\over 4 a^2} \big[ (r_+  + r_+')^2 -\Delta^2\big]^{1\over 2} \big[ (r_-  +  r_-')^2 -\Delta^2\big]^{1\over 2}\,.\label{zmidents}
 \end{eqnarray}
From which one obtains:
\begin{equation}
\sinh U ~=~ {1 \over 2\, a\, |m|} \, \Delta \, (\cos \theta + \cos \theta')   \,,
\label{geomL}
 \end{equation}
and
\begin{equation}
{z \over \Lambda} ~=~   -{1 \over 2 \,|m|} {1  \over (\cosh  U - \cos T) }~=~ -{a \over \Delta } {1 \over (\cos \theta + \cos \theta') }\,  {\sinh U \over (\cosh  U - \cos T) }   \,.
\label{zLgeom}
 \end{equation}
Thus we have recovered the core part of the Page Green function.  

We have not found any particularly simple form for $a_2 + i b_2$ or $\nu$ in terms of $U$ and $T$, but as we will now see, these functions play an essential role in avoiding jump discontinuities across critical and image surfaces.  
 
%%%%%%%%%%%%%%%%%%%
\subsubsection{Residues and pole structure}
%%%%%%%%%%%%%%%%%%%
 
It follows from (\ref{Lambdanice})  that $|z_+  z_- | =1$ and so if $z_\pm$ is inside the unit circle, then   $z_\mp$ is outside the circle.
One can also verify that
\begin{equation}
\nu ~=~  -(1+x) (u -v z) - y (v -u z) e^{-i \chi} \,.
 \label{nice1}
 \end{equation}
and
\begin{equation}
(a_2 + i b_2)  ~=~  ((1+x)^2 -y^2)^{-1}(\nu^2 + y \, e^{-i \chi} \Lambda) ~=~  (\cos \theta + \cos \theta')^{-2}(\nu^2 + y \, e^{-i \chi} \Lambda)\,.
 \label{nice2}
 \end{equation}
The importance of this identity is to show that there is a uniform, analytic choice of square root on the locus $\Lambda =0$.  More to the point, the roots $\Lambda=0$ correspond to:
\begin{equation}
z ~=~  z_\pm \quad \Leftrightarrow \quad  \nu ~=~  \pm (\cos \theta + \cos \theta')\, \sqrt{ a_2 + i b_2}  \,.
 \label{rootmatch}
 \end{equation}
Thus, we have:
\begin{equation}
{\rm Res} \Big[ {  \nu \over a\, \Lambda \sqrt{a_2 + i b_2} } \Big]_{z = z_\pm} ~=~    {(\cos \theta + \cos \theta') \over a\,  m\,(z_+ - z_-)}\,.
 \label{Residue}
 \end{equation}
Note the absence of $\pm$ signs in the result.

The factor of $ { \nu \over  \sqrt{a_2 + i b_2} }$ thus has two very important effects on the result.  First, it cancels the $(\cos \theta + \cos \theta')$ in (\ref{zLgeom}) to give the Page Green function.  If this term were not canceled then there would be the spurious ``image'' poles, alluded to earlier,  at $\theta = \pi - \theta'$. The second effect is more subtle: there is an extremely important effect on signs coming from  (\ref{rootmatch}).  Indeed, the effect is precisely the difference between:
\begin{equation}
{1 \over 2 \pi i} \oint_C { dz \over (z-z_0)(z+z_0)} ~=~ \pm {1 \over 2 z_0} \,,  \qquad
{1 \over 2 \pi i} \oint_C { dz \over (z-z_0)(z+z_0)}\,{ z \over z_0}  ~=~ {1 \over 2 z_0}   \,,
 \label{cints}
 \end{equation}
where the sign depends whether the (counterclockwise)  contour, $C$, surrounds $\pm z_0$.  The sign of the second integral does not depend upon whether the contour surrounds either $+z_0$ or $-z_0$.  This removes jump discontinuities in Green function. 
  
%%%%%%%%%%%%%%%%%%%
\subsection{Solutions and their asymptotic behavior}
\label{Gasymptotics}
%%%%%%%%%%%%%%%%%%%

In this section we set up our conventions for constructing the wiggling solutions from source currents and then examine the asymptotic behavior of these solutions in various limits.  These limits provide the crucial physical data for the regularity conditions on the wiggling supertube and in the bubble equations.

%%%%%%%%%%%%%%%%%%%
\subsubsection{Short-distance behaviour}
\label{Gphyspole}
%%%%%%%%%%%%%%%%%%%

First, in the limit that $\vec r \to \,{\vec r}{\,}'$ and $\psi \to \psi'$ one has, from  (\ref{TUdefns}), $T, U \to 0$ and
\begin{equation}
{z \over \Lambda }~\sim~  -{a \, \cos \theta \over \Delta } {\sinh U \over (\cosh  U - \cos T) }   \,, \qquad {\nu \over \sqrt{a_2 + i b_2} }~\sim~ -  { 2  \over \sqrt{\cos^2 \theta} }\,.
\label{GPartsCoincidence}
 \end{equation}
Thus
\begin{equation}
\widehat G ~\sim~   {1  \over \Delta } \,{ \cos \theta \over | \cos \theta | }\, {\sinh U \over (\cosh  U - \cos T) } ~\sim~   {1 \over \Delta }  \, {|\sinh U| \over (\cosh  U - \cos T) }  ~\sim~   {2   \over \Delta }  \, {| U| \over ( U^2 +   T^2) } \,.
\label{GCoincidenceA}
 \end{equation}
The absolute values come from the $\sqrt{\cos^2 \theta} $ and have the effect of keeping the coefficient of the singular part of the Green function positive as one crosses the surface $V=0$.

For a small separation in the GH base, the infinitesimal proper length is given by:
\begin{equation}
(\delta s)^2 ~=~ V^{-1}\big((\psi' - \psi) + q (\cos \theta_+ -  \cos \theta_-)(\phi - \phi') \big)^2 ~+~ V\, \Delta \,.
\label{proplength}
 \end{equation}
One then finds that (\ref{GCoincidenceA}) can be written as
\begin{equation}
\widehat G  ~\sim~   {4 \, q  \over | \delta s |^2 }   \,.
\label{GCoincidenceB}
 \end{equation}
Note the absolute values in the denominator.  As we will discuss below, this result means that $\widehat G$ is the properly normalized Green function for our purposes.

The Green function, $\widehat G$,  depends upon $\psi, \psi'$ via $\cos T$, where $T$ is given by (\ref{TUdefns}) and so $\widehat G$ is periodic on the interval $[0,4 \pi q]$.  It is, of course, elementary to produce a Green function with a reduced periodicity, like $4 \pi$, simply by summing over image sources appropriately.   Here we will work with the function $\widehat G$ and sources that have the same periodicity and so the solution to the Laplacian with a source density, $\rho (\psi')$, along the GH fiber located at  ${\vec r}{\,}'$ is therefore given by:
\begin{equation}
\Phi(\vec r, \psi; \,{\vec r}{\,}')  ~=~  \int_0^{4\pi q}  \, \widehat G(\vec r, \psi; \,  {\vec r}{\,}', \psi')\,  \rho (\psi')\ d \psi'  \,.
\label{FundSol}
 \end{equation}
 It is convenient to define the associated total charge by:
\begin{equation}
Q ~\equiv~ \int_0^{4\pi q}  \,  \rho (\psi')\ d \psi'  \,.
\label{TotChg}
 \end{equation}
In the coincidence limit, when one approaches the source,  $\vec r \to \,{\vec r}{\,}', \psi \to \psi'$ the dominant contribution to (\ref{FundSol}) comes from integrating through the singularity and so
\begin{eqnarray}
\Phi &\sim&  \int_{\psi-\epsilon}^{\psi+\epsilon}   {4   q  \over  \big|V^{-1}  (\psi - \psi')^2 ~+~ V\, \Delta \big|}    \rho (\psi')\ d \psi'  \nonumber \\
&\sim&     {4  q  \over  \Delta}    \rho (\psi)  \arctan\Big({\epsilon \over |V| \Delta}\Big) ~\to~{4\pi    q  \over  \Delta}\,     \rho (\psi)   \,.
\label{PhiLim}
 \end{eqnarray}
This is precisely what one should expect:  For a constant charge, the charge density is $\rho = {Q \over 4 \pi q}$ and thus
(\ref{PhiLim}) gives $\Phi \sim {Q   \over \Delta}$, which is the canonical form in the GH base.   Moreover, if one approaches any line distribution of charge then is should appear, locally, like a constant charge and hence one must obtain (\ref{PhiLim}).  This is the reason for selecting the normalization of $\widehat G$ in (\ref{Gnice}).

%%%%%%%%%%%%%%%%%%%
\subsubsection{Asymptotics at the Gibbons-Hawking points}
\label{GatGHpts}
%%%%%%%%%%%%%%%%%%%
 
Another important limit is  to understand the behavior of $\widehat G$ and the solution $\Phi$, defined in (\ref{FundSol}), near the GH points.  This evaluation is straightforward and one finds that at $r_\pm = 0$, or $\vec r =(0,0,\pm a)$, one has
\begin{equation}
\widehat G(\vec r =(0,0,\pm a) , \psi; \,  {\vec r}{\,}', \psi') ~=~ { 1 \over r_\pm'}\,.
\label{GHlims}
 \end{equation}
Observe that this is independent of $\psi$ and $\psi'$ and so the solution (\ref{FundSol}) depends only on the total charge of the supertube, and not on the way this charge is distributed on the supertube world-volume: 
\begin{equation}
\Phi(\vec r =(0,0,\pm a) , \psi; \,{\vec r}{\,}',\psi' )  ~=~  { 1 \over r_\pm'} \, \int_0^{4\pi q}  \,  \rho (\psi')\ d \psi' ~\equiv~ { Q \over r_\pm'}\,.
\label{FundSolGHvals}
 \end{equation}
One should have anticipated a result of the form (\ref{FundSolGHvals}) from the outset:  At the GH points the GH fiber, parametrized by $\psi$ collapses to a point but the space-time looks locally smooth.  Regularity therefore requires that any non-trivial Fourier mode on the fiber must vanish at the GH point and only the constant mode can survive.  Thus the GH points only feel the total charge of the fluctuating solution.

Now recall that the bubble equations at the GH points are local conditions and so it follows that in a geometry with fluctuating electric charge densities along the GH fiber, the bubble equations {\it for the GH points} are unchanged by the fluctuations and only depend upon the total charge of the fluctuating component.

%%%%%%%%%%%%%%%%%%%
\subsubsection{Behaviour at infinity}
%%%%%%%%%%%%%%%%%%%

We now consider the behavior of the Green functions and solutions of the form (\ref{FundSol}) as $r = |\vec r|$ becomes large.  This corresponds to taking $\xi$ to be large and (\ref{coordsa}) implies
\begin{equation}
 e^{-2 \xi}  ~\sim~   {a \over 2\, r} \,.
  \label{infasymp}
 \end{equation}
One can now take the limit of large $\xi$ in (\ref{Gnice}) but it is simpler to take the large $\xi$ limit directly in  (\ref{int1}).  In particular, note that  in the large $\xi$ limit, with the analytic continuation $\tau - \tau' = i \lambda$, one has:
\begin{eqnarray}
\zeta   &\sim&   \coeff{1}{2}\, e^\xi (\cosh \xi'  \cos (\tau - \tau')  ~-~    \sinh \xi'  \cos (\varphi_1 -\varphi_1') ) \\
&=&    \coeff{1}{2}\, e^\xi (a_1 \cosh \lambda   ~+~    i b_1 \sinh \lambda   )\,.
 \label{zetaasymp}
 \end{eqnarray}
Thus
\begin{eqnarray}
\widehat G &\equiv&  {1 \over \sqrt{2} \, a}\,  \int_{-\infty}^{\infty}     { \zeta \over (2 \zeta ^2 - (1+  \cos \gamma) )^{3\over 2} } \,d \lambda  \\
&\sim&  \int_{-\infty}^{\infty}     {1 \over 4\, a\, \zeta^2}\,  d \lambda  ~=~{1 \over 4\, a} \int_{-\infty}^{\infty}     { 1 \over  (a_1 \cosh \lambda   +  i b_1 \sinh \lambda)^2 } \,d \lambda \\
&=&{ 1 \over 2\,a\, (a_1  -   i b_1)   (a_1  +    i b_1)   } ~\sim~{ 2\,   e^{-2 \xi} \over  a\, (\cosh \xi'  -   z \sinh \xi')   (\cosh \xi'  -   z^{-1} \sinh \xi')  }  \\
&\sim& -  {1 \over r}{ z    \over \sinh \xi'  \cosh \xi'( z  -   \coth \xi')   ( z  -   \tanh \xi') }  \,.
\label{Ginfty}
 \end{eqnarray}

Now suppose that a charge density has the Fourier series:
\begin{equation}
 \rho(\psi')  ~=~   \sum_{n=-\infty}^\infty \alpha_n \, e^{{i n \over 2 q} \psi'}\,,
  \label{exchgdens}
 \end{equation}
for which reality implies $\alpha_{-n} = \overline{\alpha_n}$.  Then one finds that the asymptotic form of the fundamental solution, (\ref{FundSol}), is given by:
\begin{equation}
\Phi(\xi', \psi)  ~\sim~   {2  i q \over r}\, \sum_{n=-\infty}^\infty   \alpha_n e^{{i n \over 2 q} \psi}  \oint_{|z|=1}   \, { z^{-n}  \, dz   \over \sinh \xi'  \cosh \xi'( z  -   \coth \xi')   ( z  -   \tanh \xi') } \ d \psi'  \,.
\label{FundSolasymp1}
 \end{equation}
 This integral is an elementary residue for $n$ non-positive, while
 the result for $n$ positive is the complex conjugate of the result
 for $n$ negative.  The solution is thus:
\begin{equation}
\Phi(\xi', \psi)  ~\sim~   {4 \pi q \over r}\, \Big[ \alpha_0 ~+~ \sum_{n=1}^\infty   \big(\alpha_n  w^n ~+~  \bar \alpha_n  \bar w^n   \big) \Big] \,, \qquad w ~\equiv~ \tanh \xi' \, e^{{i  \over 2 q} \psi} \,.
\label{FundSolasymp2}
 \end{equation}
 Observe that we have the following limits
\begin{equation}
\Phi(0, \psi)  ~\sim~   {4 \pi q \over r}  \alpha_0\,, \qquad  \Phi(\xi'= \infty, \psi)  ~\sim~  {4 \pi q \over r}  \rho(\psi) \,,
\label{FundSolasymplims}
 \end{equation}
 and $\xi'=0$ puts the supertube on the axis between the two GH
 points, while $\xi' \to \infty$ corresponds to moving the supertube
 off to infinity in the GH space.  Thus the asymptotics of the
 solution averages out if the supertube is between the two GH points
 and is directly proportional to the density if the supertube is at
 infinity.

We are generically interested in putting bounds upon asymptotic values
of our solutions, and to that end we note that the asymptotic
functional dependence (\ref{FundSolasymp2}) satisfies the Laplace
equation on the semi-infinite cylinder defined by $-\log(\tanh(\xi'))$
and $\psi$.  It follows from the maximum modulus theorem that the
asymptotic behaviour of the fundamental solution as $\xi \to \infty$
is bounded by the boundary values, (\ref{FundSolasymplims}).

%%%%%%%%%%%%%%%%%%%%%%%%%%%%%%%%%%%%%
\section{The fully back-reacted solution for a wiggling supertube}
\label{sec:fullsol}
%%%%%%%%%%%%%%%%%%%%%%%%%%%%%%%%%%%%%

It now remains to assemble all the constituents of the fully back-reacted solution and to check the regularity at the supertube and at the GH points and then impose any constraints required by regularity and absence of CTC's.  We  found, in Section \ref{GatGHpts},  that in a geometry with fluctuating electric charge densities along the GH fibers, the bubble equations are unchanged by the fluctuations and only depend upon the total charge of the fluctuating component.  This has the important consequence that regularity {\it at the GH points}  for the corresponding wiggling supertube has exactly the same form as the regularity conditions for the round, non-wiggling supertube. There are, however, two important differences.  

\begin{figure}
\label{Pic1}
\centering
 \includegraphics[width=0.3\linewidth]{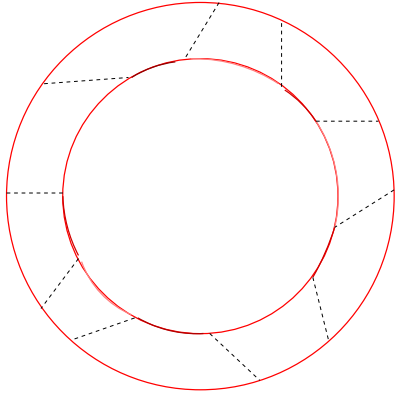}
\caption{\footnotesize A round supertube can retain its round shape but concentrate or dilute charges in different regions as it expands or contracts. Some large sectors of the original tube can be mapped into smaller sectors, concentrating charge, while some sectors can become expanded resulting in a charge dilution.   Even though the shape of the supertube remains circular, the charge density modes induce a radius change and can thus be seen as special class of shape mode.}
\end{figure}

The first difference is that around the wiggling supertube the standard regularity conditions are replaced by a simple generalization to a local regularity conditions along the fiber.  Secondly, we find that the dipole-dipole interactions will allow large negative values of the angular momentum, $\widehat J$, of the supertube, and thus enable entropy enhancement, but this also has the concomitant effect of having the supertube approach the critical ($V=0$) surface where it becomes larger and floppier, while remaining round.  The fluctuating charge density thus generates a displacement in the base space and an associated change of the supertube radius.  We can then interpret the result as a shape mode that preserves roundness:  The fluctuations in charge density can be thought of as being the result of differential expansion of different parts of the supertube (while keeping it round) diluting charge in some regions and concentrating it in others.  This process is depicted in Fig. \ref{Pic1}.

We now construct the solution in detail and exhibit all the regularity constraints.

%%%%%%%%%%%%%%%%%%%%%%%%%%%%%%%%%%%%%
%%%%%%%%%%%%%%%%%%%%%%%%%%%%%%%%%%%%%

%%%%%%%%%%%%%%%%%%%
\subsection{The constituents of the solution}
%%%%%%%%%%%%%%%%%%%
As shown in Section \ref{sec:solution}, the full solution for a round supertube with fluctuating charge densities inside a GH space is given by 8 harmonic functions in five dimension:
\begin{eqnarray}
V &=&  q\, \Big({ 1 \over r_+} ~-~ {1 \over r_-} \Big)\,, \qquad  K^I ~=~
k_I\, \Big({ 1 \over r_+} ~+~ {1 \over r_-} \Big) ~+~ \delta^I_3 \, {\hat k \over \Delta}  \,, \nonumber \\
\qquad  L_I &=& \ell_I \Big({ 1 \over r_+} ~-~ {1 \over r_-} \Big)~+~ \sum_{A=1}^2 \delta^A_I \lambda_A (\vec r, \psi;  {\vec r}{\,}' )  \,, \label{wigglyharms} \\
\qquad  M &=&   m_0 +~ m \Big({ 1 \over r_+} ~+~  {1 \over r_-}\Big)~+~  j (\vec r, \psi;  {\vec r}{\,}' )  \,,
\nonumber
\end{eqnarray}
where the components of the solution sourced by the fluctuating densities are
\begin{eqnarray}
\lambda_A(\vec r, \psi; \,{\vec r}{\,}')  ~=~  \int_0^{4\pi q}  \, \widehat G(\vec r, \psi; \,  {\vec r}{\,}', \psi')\,  \rho_A (\psi')\ d \psi'  \,, \nonumber \\
  j(\vec r, \psi; \,{\vec r}{\,}')  ~=~  \int_0^{4\pi q}  \, \widehat G(\vec r, \psi; \,  {\vec r}{\,}', \psi')\,  \hat \rho (\psi')\ d \psi'   \,,
\label{wigglybits}
\end{eqnarray}
and
\begin{equation}
\ell_I    ~=~  -\frac{1}{2 q} C_{IJK} k_J k_K\,, \qquad m    ~=~  \frac{1}{12 q^2} C_{IJK} k_I k_J k_K\,.
\label{lmparamers}
\end{equation}
Note that in (\ref{wigglyharms}) we have chosen a gauge so that the $K^I$ have equal charges at the GH points.  For simplicity, we will also assume that $q >0$ so that integrals like (\ref{wigglybits}) run in the canonical (positive) orientation.  We will also assume that the $k_I$ are all positive so that it is simple to assign signs to square roots in the metric and other functions.

With these choices we have:
\begin{eqnarray}
Z_1 &=& - {4\, k_2 k_3 \over q}  \, { 1 \over (r_+ - r_- )}~+~  {  k_2 \hat k  \over q}  \, { F \over \Delta}~+~ \lambda_1\,,  \\
Z_2 &=&  - {4\, k_1 k_3 \over q}  \, { 1 \over (r_+ - r_- )}~+~  {  k_1  \hat k  \over q}  \, { F \over \Delta}~+~ \lambda_2\,, \qquad Z_3 ~=~  - {4\, k_1 k_2 \over q}  \, { 1 \over (r_+ - r_- )} \,,\\
\mu &=&   m_0 + j +  {4\, k_1 k_2 k_3 \over q^2}  \,
{  (r_+ + r_- ) \over (r_+ - r_- )^2} +    { k_1 k_2 \hat k \over q^2 \,\Delta}  (F^2 - \coeff{1}{2}) +  { 1 \over 2q }\,  (k_1 \lambda_1 + k_2 \lambda_2 ) F\,,
\label{Zmufns}
\end{eqnarray}
where 
\begin{equation}
F   ~\equiv~  -\frac{(r_+ + r_-)}{(r_+ -  r_-)}  \,.
\label{Fdefn}
\end{equation}

It is also convenient to define a special value, $\widehat F$, of $F$ at the supertube:
\begin{equation}
\widehat F   ~\equiv~ F_{\vec r = \vec r'} ~=~ -\frac{(r'_+ + r'_-)}{(r'_+ -  r'_-)}  \,,
\label{hatFdefn}
\end{equation}
where $r'_\pm ~=~ |\vec r' - \vec r_\pm|$.

Note that we do not have an explicit solution to the (linear) equations, (\ref{Lorgauge}) and (\ref{omegaeqn}), for $\vec \omega$, but we will not need the explicit solution in the following.

%%%%%%%%%%%%%%%%%%%
\subsection{Supertube regularity}
\label{STregular}
%%%%%%%%%%%%%%%%%%%

Consider a supertube located at $\vec r = {\vec r}{\,}' $ in the
$\IR^3$ base of the GH space and with charges associated with $K^3$,
$L_1$, $L_2$ and $M$.  That is, suppose that only these functions
diverge as $\Delta^{-1}$ for $\Delta \equiv | \vec r - {\vec r}{\,}' |
\to 0$.  Moreover, as above, we allow $L_1$, $L_2$ and $M$ to have
densities that depend upon $\psi$.  The regularity of this supertube
amounts to requiring that the metric (\ref{D1D5Pmetric}) be regular
and without CTC's at the supertube.  There are two potential dangers:
(i) Divergences parallel to the fiber, and (ii) Dirac strings in $\vec
\omega$. As discussed in \cite{Bena:2008dw}, for this supertube, the
former condition requires that for all $\psi$:
\begin{equation}
\lim_{\Delta  \to 0}\,  \Delta^2 \left[ Z_3\, (K^3)^2 ~-~ 2 \mu V K^3 ~+~ Z_1 Z_2 V \right] ~=~ 0 \,,
\label{regconda}
\end{equation}
where $\Delta =| \vec r - {\vec r}{\,}' |$.

To analyze Dirac strings, consider any two-sphere, $S^2$, in the
$\IR^3$ base of the GH space.  If the magnetic field, $\vec \nabla
\times \vec \omega$, has a monopole source inside the $S^2$ then $\vec
\omega$ will have a Dirac string somewhere on this $S^2$.  Let $\phi$
be the polar angle around this Dirac string and suppose that the
charge density is independent of $\psi$, then one can let $\psi$ vary
in such a way that $(d\psi +A)$ vanishes as $\phi$ goes around the
string.  If one does this, the closed curve around the Dirac string is
necessarily time-like in the metric (\ref{D1D5Pmetric}). If the charge
density varies then the foregoing argument may not work, but if the
variation is small then there will still be CTC's associated with the
Dirac string.  It is conceivable that one might find a conspiracy for
large variation in the charge density and that one might avoid CTC's
associated with Dirac strings, but it is simpler to require that there
are no Dirac strings in $\vec \omega$ no matter how it depends upon
$\psi$. That is, for all $\psi$ we impose the condition:
\begin{equation}
\int_{S^2}\, (\vec \nabla \times \vec \omega)\cdot \hat n ~=~ 0 \,,
\label{nomonoplesA}
\end{equation}
where the $S^2$ is any small sphere in the $\IR^3$ base drawn around the charge source and $\hat n$ is the unit outward normal to this $S^2$.

Suppose that the charge density at $\vec r ={\vec r}{\,}' $ has a
monopolar magnetic field and consider a vanishingly small $S^2$ around
${\vec r}{\,}' $.  As $\Delta \to 0$, the Cartesian $\IR^3$ components
of $\vec \omega$ will diverge, in a direction-dependent manner, as
$\Delta^{-1}$.  Also assume that ${\vec r}{\,}' $ is not a GH point
and so $V$ and $A \equiv \vec A\cdot d\vec y$ are smooth at $\vec r
={\vec r}{\,}' $. Now imagine integrating equation (\ref{covkeqn})
over this $S^2$: The terms involving $\partial_\psi\vec \omega$ and
$\vec A \times \partial_\psi\vec \omega$ and $V \vec A \partial_\psi
\mu$ do not contribute to this integral as $\Delta \to 0$. Therefore
we find that
\begin{equation}
\int_{S^2}\, (\vec \nabla \times \vec \omega)\cdot \hat n ~=~   \int_{S^2}\,   V\, \Big(\vec \nabla \mu ~-~ \sum_{I=1}^3 \, Z_I \, \vec \nabla \big(V^{-1} K^I  \big)\Big)\cdot \hat n    \,.
\label{nomonoplesB}
\end{equation}
For this to vanish as $\Delta \to 0$ one must have:
\begin{equation}
\lim_{\Delta \to 0}\,  \Delta  \big[ V  \mu ~-~Z_3\,  K^3 \big] ~=~ 0
.\label{regcondb}
\end{equation}
This result is of the same form as the result for constant charge densities, but here this equation is to be applied as a functional constraint that must be true for all $\psi$, just like (\ref{regconda}).

One can use (\ref{regcondb}) in (\ref{regconda}) to obtain the simpler condition:
\begin{equation}
\lim_{\Delta \to 0}\,  \Delta^2 \big[  V Z_1 Z_2  ~-~ Z_3 \, (K^3)^2  \big] ~=~ 0  \,.
\label{regcondc}
\end{equation}
It is slightly more convenient to work with  (\ref{regcondb}) in (\ref{regcondc}).

Let $\rho_1(\psi)$, $\rho_1(\psi)$ and $\hat \rho (\psi)$ denote the charge densities that generate the functions $L_1$, $L_2$ and $M$ via the expression (\ref{FundSol}).  Let $\hat k$ be the magnetic dipole charge of the supertube, that is:
\begin{equation}
K^3 ~\sim~ {\hat k \over \Delta } \,,
\label{magchg}
\end{equation}
as $\Delta \to 0$.  In addition, let  $\widehat V$, $\widehat K^1 $, $\widehat K^2 $, $\widehat  L_3 $ and  $\widehat  Z_3 $ denote the (finite, $\psi$-independent) values of the corresponding functions at the supertube location ($\vec r ={\vec r}{\,}' $).  Using the limiting behavior in (\ref{PhiLim}), one finds that (\ref{regcondb}) becomes
\begin{equation}
4 \pi q \big[ \, \widehat K^1 \rho_1(\psi) ~+~\widehat K^2 \rho_2(\psi)~+~ 2\, \widehat V \hat \rho (\psi) \, \big] ~=~ \hat k \, \widehat L_3  \,.
\label{densrelna}
\end{equation}
Similarly, (\ref{regcondc}) becomes
\begin{equation}
\big[ \,4 \pi q  \rho_1(\psi) ~+~\hat k \widehat V^{-1} \widehat K^2  \, \big] \,\big[ \,4 \pi q  \rho_2(\psi) ~+~\hat k \widehat V^{-1} \widehat K^1  \, \big] ~=~ \hat k^2  \widehat V^{-1}  \widehat Z_3  \,.
\label{densrelnb}
\end{equation}
Combining this with  (\ref{densrelna}) gives the simple relationship between the densities:
\begin{equation}
 \hat k   \hat \rho (\psi)  ~=~  2 \pi q \,   \rho_1(\psi)  \,  \rho_2(\psi)   \,.
\label{densrelnc}
\end{equation}

Note that if one specifies $\rho_1(\psi)$ then these identities fix $\rho_2(\psi)$ and $\hat \rho(\psi)$ in terms of $\psi$-independent geometric factors.

Finally, define the charges:
\begin{equation}
\widehat Q_A ~\equiv~ \int_0^{4\pi q}  \,  \rho_A (\psi')\ d \psi'  \,, \qquad \widehat J ~\equiv~ \int_0^{4\pi q}  \,  \hat \rho(\psi')\ d \psi'  \,.
\label{STchgs}
\end{equation}
It is useful to note that  (\ref{densrelna}) integrates to:
\begin{equation}
\big[ \, \widehat K^1 \widehat Q_1 ~+~\widehat K^2 \widehat Q_2 ~+~ 2\, \widehat V \widehat J\, \big] ~=~ \hat k \, \widehat L_3  \,.
\label{chgrelns}
\end{equation}
For constant charge densities one thus gets this equation combined with  (\ref{densrelnc}), which yields
\begin{equation}
  \widehat J ~=~  \frac{\widehat Q_1 \, \widehat Q_2 }{2 \hat k }  \,.
\label{angmomrelna}
\end{equation}
%

%%%%%%%%%%%%%%%%%%%
\subsection{The bubble equations}
\label{BubbEqns}
%%%%%%%%%%%%%%%%%%%

The bubble equations at the two GH points are given by setting $\mu=0$ at each point:
\begin{equation}
 \mu_\pm  ~=~ m_0 ~+~ \frac{ 2\,k_1 k_2 k_3 }{ a q^2}  ~+~  \Big(  \frac{ \hat k k_1 k_2 }{2 q^2} ~\pm~
 \frac{ 1 }{2 q} \,\sum_{A=1}^2 k_A \widehat Q_A  ~+~  \widehat J \Big) \, {1 \over r_\pm'}   ~=~ 0\,.
\label{bubbeqna}
\end{equation}
The difference of these two equations gives:
\begin{equation}
\mu_+ - \mu_-   ~=~  \bigg(  \frac{ \hat k k_1 k_2 }{2 q^2}  ~+~  \widehat J \bigg) \, \Big({1 \over r_+'}  -  {1 \over r_-'} \Big)~+~
  \frac{ 1 }{2 q} \, \bigg( \sum_{A=1}^2 k_A \widehat Q_A   \bigg) \, \Big({1 \over r_+'}  +  {1 \over r_-'} \Big) ~=~ 0\,.
\label{bubbdiffa}
\end{equation}
This is precisely equation (\ref{chgrelns}) and this observation is the analog of the fact that, when the charge densities are constant, the sum of the bubble equations vanishes.   One can write  (\ref{bubbdiffa}) as
\begin{equation}
\widehat F  ~=~  -2 q \, \bigg( \sum_{A=1}^2 k_A \widehat Q_A   \bigg)^{-1} \bigg(  \frac{ \hat k k_1 k_2 }{2 q^2}  ~+~ \widehat J \bigg)  \,.
\label{bubbdiffb}
\end{equation}
This equation determines the scale-invariant ratio, $r_+'/r_-'$  while the remaining bubble equation in (\ref{bubbeqna}) determines $m_0$.

Thus, once one specifies all the charges, the geometry is fixed.  The non-trivial constraints come from the fact that the charge densities on the supertube are all related and thus the charges $\widehat Q_A$ and $\widehat J$ must be related.  Indeed, (\ref{densrelnb}) gives:
\begin{equation}
4 \pi q  \rho_2(\psi)  ~=~ -\frac{\hat k k_1}{q}   \bigg[ \,4 \pi q  \rho_1(\psi)  ~+~\frac{\hat k k_2}{q} \widehat F   \, \bigg]^{-1} \, \bigg[ \,4 \pi q  \rho_1(\psi)  \widehat F ~+~\frac{\hat k k_2}{q}   \, \bigg]    \,.
\label{densrelnd}
\end{equation}
One then gets $ \hat \rho (\psi)$ from (\ref{densrelnc}).   In particular, from (\ref{STchgs}) one has
\begin{equation}
 \widehat J ~=~  -\frac{ k_1}{2\,q} \int_0^{4\pi q}  \,    \rho_1(\psi)     \bigg[ \,4 \pi q  \rho_1(\psi)  ~+~\frac{\hat k k_2}{q} \widehat F   \, \bigg]^{-1} \, \bigg[ \,4 \pi q  \rho_1(\psi)  \widehat F ~+~\frac{\hat k k_2}{q}  \, \bigg]   \ d \psi \,.
\label{Jhatres}
\end{equation}
%

%%%%%%%%%%%%%%%%%%%%%%%%%%%%%%%%%%%%%
\section{Comparison with probe supertubes}
\label{sec:DBI}
%%%%%%%%%%%%%%%%%%%%%%%%%%%%%%%%%%%%%

In the previous section we obtained the regularity conditions near the wiggling supertube in the fully back-reacted solution with two GH points and a supertube.  Obviously one would like to generalize this solution to include more supertubes and more GH points but, as we discussed earlier, such fully back-reacted solutions are, at present, beyond the limit of explicit computation and so we must seek another approach to this problem.

It is natural to fall back upon a probe approximation and use the DBI action that describes supertubes in the regime of parameters in which they do not back-react on the geometry. This action can be used to analyze more general configurations, but the 
problem then becomes one of determining the extent to which the DBI action captures the correct physical behavior of the complete physical (back-reacted) solution. In \cite{Bena:2008dw}, it was shown that for supertubes with constant charge densities, the DBI action of supertubes always precisely captures the conditions necessary for smoothness and regularity of the fully back-reacted metric at the supertube location. Our purpose here is to generalize this to supertubes that have non-trivial density modes, and show that the solutions of the DBI action always correspond to supertubes whose back-reacted solution is smooth and free of CTC's. We show that not only can one use the DBI action to extract the local regularity around the supertube but one can combine the information coming from this action with the general analysis of solutions in section \ref{sec:solution} to infer the full regularity conditions at the GH points that lie far from the supertube, and thus find the complete set of functional bubble equations that govern multi-center solutions containing one wiggly supertube in an arbitrary Gibbons-Hawking base space. 

%%%%%%%%%%%%%%%%%%%
\subsection{Regularity conditions from the supertube Born-Infeld action   }
%%%%%%%%%%%%%%%%%%%

The Born-Infeld  analysis of  supertubes is best  done in  the duality frame  where the  supertube  has D0  and  F1 electric  charges and  D2 magnetic dipole  charges, and  for this we  first have to  dualize the background given in (\ref{11Dmetric}) and (\ref{11Dthreeform})  to this type  IIA frame.  The details of this transformation may be found in  \cite{Bena:2008dw} and here we will also adopt the  same
notation and conventions. 

We first reduce to ten dimensions along  $x_5$ and then do two T-dualities along $x_7$ and $x_8$. The F1 string and the dipole D2 brane are wrapped along the $x_6=z$. The resulting metric (in the string frame) is:
\begin{equation}
ds^2_{10} =  - \displaystyle \frac{1}{\sqrt{Z_2Z_3}Z_1} \, (dt + k)^2 ~+~
\sqrt{Z_2Z_3}\, ds_4^2 ~+~  \frac{\sqrt{Z_2Z_3}}{Z_1}\, dz^2 ~+~ \sqrt{\frac{Z_2}{Z_3}}\, ds^2_{T^4}
\end{equation}
and the dilaton and the Kalb-Ramond field are  given by:
\begin{equation}
e^{\Phi} = \left( \ds\frac{Z_2^3}{Z_3Z_1^2}\right)^{\frac{1}{4}} ~, \qquad B = (Z_1^{-1} - 1) dt \wedge dz + Z_1^{-1} k
\wedge dz  \,.
\end{equation}
The non-trivial RR potentials are:
\begin{eqnarray}
C^{(1)} &=& (Z_2^{-1} - 1) dt + Z_2^{-1}k~, \\
C^{(3)} &=& \big(\zeta_a ~+~  V^{-1} K^1 \xi^{(2)}_a  \big) \,
\Omega_-^{(a)}  \wedge dz  - \big(Z_1^{-1} ( dt +k) \wedge B^{(2)}
~+~ dt \wedge A^{(1)}  \big) \wedge dz \,, \label{C3IIA}
\end{eqnarray}
where $\Omega_-^{(a)}$, $\xi^{(2)}$, $k$ and $Z_I$ are given respectively by (\ref{twoforms}),  \eqref{vecpotdefns}, \eqref{kansatz} and \eqref{ZIform}, and
\bea
 \vec{\nabla}\times\vec{\zeta} = -\vec{\nabla} L_3 \,.
\eea

We will denote the world-volume coordinates on the supertube by
$\xi^0$, $\xi^1 $ and $\xi^2\equiv \theta$.  To make the supertube
wrap $z=x_6$ we take $\xi^1=z$ and we will fix a gauge in which $
\xi^0=t$.  Note that $z\in (0,2\pi L_z)$. The profile of the tube,
parameterized by $\theta$, lies in the four-dimensional non-compact
Gibbons-Hawking space. We will here take the profile to be round,
$\theta=k \psi$,  with dipole charge $k$, but will allow for density
modes: the gauge field, $ \mathcal{F}$, living on the world-volume of
the supertube will depend on $\theta$. We take
\begin{equation}
2\pi\alpha' F \equiv \mathcal{F} = \mathcal{F}_{tz} dt\wedge dz +
\mathcal{F}_{z\theta } dz\wedge d\theta \,.
\end{equation}
Supersymmetry requires that $\mathcal{F} _{tz} =1$ \cite{Mateos:2001qs}, but $\mathcal{F}_{z\theta}$, which corresponds to the D0-charge density, can be an arbitrary function, of $\theta$, $\mathcal{F}_{z\theta}=\mathcal{F}_{z\theta}(\theta)$.

The supertube action is a sum of the DBI and Wess-Zumino (WZ) actions:
\begin{equation}
S =  - T_{D2} \int d^3\xi e^{-\Phi}\sqrt{ - \text{det}
\left(\widetilde{G}_{ab} + \widetilde{B}_{ab} + \mathcal{F}_{ab}
\right)} + T_{D2} \int d^3\xi [\widetilde{C}^{(3)} +
\widetilde{C}^{(1)}\wedge (\mathcal{F} + \widetilde{B})] \,,
\label{DBIWZ}
\end{equation}
where $\widetilde{G}_{ab}$ and $\widetilde{B}_{ab}$ are
the induced metric and Kalb-Ramond field on the world-volume of the tube. We have also chosen the
orientation such that $\epsilon_{tz\theta}=1$, and throughout this section all the functions will be evaluated at the location of the supertube, and we will omit the hats that we used to note this in section \ref{sec:fullsol}

After some algebra, the action simplifies to:
\begin{multline}
S = T_{D2} \int d^3\xi  \,\bigg\{ \left[ \left(\ds\frac{1}{Z_2} -
1\right) \mathcal{F}_{z\theta} + \ds\frac{K^1}{Z_2V} + \left(
\ds\frac{\mu}{Z_2} - \ds\frac{K^2}{V} \right)(\mathcal{F}_{tz}-1)
\right] \\- \bigg[ \ds\frac{1}{V^2Z_2^2} \big[(K^1  -  V
(\mu(1-\mathcal{F}_{tz})- \mathcal{F}_{z\theta}) )^2 + V Z_2Z_3
(1-\mathcal{F}_{tz})(2-Z_1(1-\mathcal{F}_{tz})) \big]
\bigg]^{1/2}\bigg\}  \,.
\end{multline}
For a supersymmetric configuration ($\mathcal{F}_{tz} = 1$) we have:
\begin{equation}
S_{\mathcal{F}_{tz}=1} = S_{DBI} + S_{WZ} = -T_{D2} \int dt dz
d\theta \, \mathcal{F}_{z\theta} \,. \label{SimpAction}
\end{equation}
The foregoing supertube carries F1 and D0 ``electric" charges, given by:
\begin{equation}
N_{1}^{ST} =
\ds\frac{1}{T_{F1}} \ds\int d\theta  ~ \ds\frac{\partial
\mathcal{L}}{\partial
\mathcal{F}_{tz}}\bigg|_{\mathcal{F}_{tz}=1}\,, \qquad\qquad   N_{2}^{ST} = \ds\frac{T_{D2}}{T_{D0}} \ds\int dz d\theta ~
\mathcal{F}_{z\theta}\,.
\label{STcharges}
\end{equation}
To avoid the proliferation of parameters, it is convenient to use the system of units described in the appendix of  \cite{Bena:2008dw}.  With these conventions, the supertube supergravity charges,  $\hat{Q}_1$ and $\hat{Q}_2$, and dipole charge, $\hat{k}$, are related to the integer charges, $N_{1}^{ST}$, $N_{2}^{ST}$ and $k$ via:
\begin{equation}
\hat{Q}_A = {1\over 4} N_{A}^{ST}\,,\quad \hat{k} = {1\over 2}{q\,k}\,,
\label{supertubecharges}
\end{equation}
and the values of the tensions are:
\begin{equation}
T_{D0}=1\,,\quad 2\pi L_z\, T_{F1}=1\,,\quad {2\pi\, T_{D2}\over T_{F1}}=1\,.
\label{tensions}
\end{equation}
The Hamiltonian density is:
\begin{equation}
\mathcal{H}|_{\mathcal{F}_{tz}=1} = \int dz\,
\left[\ds\frac{\partial{\mathcal{L}}}{\partial\mathcal{F}_{tz}}
\mathcal{F}_{tz}-\mathcal{L}\right]_{\mathcal{F}_{tz}=1} \equiv~~4\left(
\frac{\rho_1}{k} + \frac{\rho_2}{k}\right)\,, \label{HamDensity}
\end{equation}
where we introduced $\rho_1$ and $\rho_2$, the charge densities of F1 and D0 charges respectively\footnote{The factor of $4$ in (\ref{HamDensity}) comes from the relation (\ref{supertubecharges}) for $\hat{Q}_A$.}:
\bea
 {\rho_1\over  k}= {1\over 4\,T_{F1}} \ds\frac{\partial
\mathcal{L}}{\partial \mathcal{F}_{tz}}\bigg|_{\mathcal{F}_{tz}=1}
 \,, \qquad {\rho_2 \over k} ={T_{D2}\over 4} \int dz\, \mathcal{F}_{z\theta} \,.
 \label{densities}
\eea
Note that, in the Born-Infeld calculation, we are treating a supertube of dipole charge $k$ as a supertube of dipole charge one wrapped $k$ times, and hence the ${\rho_I \over k}$ are densities per unit strand of the supertube, and the total density is $\rho_I$. We want to remind the reader that $\rho_1$, $\rho_2$ are not constant but vary with $\psi$.

One can easily integrate this to get the total Hamiltonian of the
supertube
\begin{equation}
\int d \theta ~ \mathcal{H}|_{\mathcal{F}_{tz}=1} =
N_1^{ST} + N_2^{ST}\,.
\end{equation}
Thus the energy of the supertube is the sum of its conserved
charges which shows that the supertube is indeed a BPS object.

In the DBI analysis, the first equation determining the regularity of the supertube comes from the relation that gives the F1 charge density $\rho_1$
\bea
\rho_1= {k\over 4 \,T_{F1}}\left.\frac{\partial{\mathcal{L}}}{\partial\mathcal{F}_{tz}}\right|_{\mathcal{F}_{tz}=1} = {k\,T_{D2}\over 4 \,T_{F1}}\left[-{K^2\over V}+{Z_3\over K^1+V \mathcal{F}_{z\theta} }\right]\,.
\eea
Using the relation (\ref{supertubecharges})  for $\hat{k}$, the values of the tensions (\ref{tensions}) and the definition of $\rho_2$ (\ref{densities}), this can be rewritten as
\bea \label{radiusrelation}
 \left({4\pi q\,\rho_1 + \hat{k} \frac{K^2}{V}}\right) \left({4\pi q\,\rho_2 + \hat{k} \frac{K^1}{V}}\right) = \hat{k}^2 \frac{Z_3}{V} \,,
\eea
and this is exactly the same equation as \eqref{densrelnb}.

\medskip

The second regularity condition comes from expressing the angular momentum of the tube as a function of the charge densities. Note that, in contrast to the supergravity construction in which this angular momentum is a free parameter,  in the DBI calculation the supertube angular momentum is related directly to the other charges. The angular momentum density $\rho$ along the $\psi$ circle is, in general, given by
\bea
{ \rho\over k} ~=~  {1\over 8} \int dz \, \ds\frac{\partial \mathcal{L}}{\partial \dot{\psi}} \,.
\eea
One should note that, even if one is interested in time-independent solutions,  in order to compute $\rho$ one has to allow a time dependence initially, compute and $\rho$, and only then impose time independence. The angular momentum of the supertube is then given by
\bea
 J = \int d\theta \rho\,.
 \label{Jsuper}
\eea
After lengthy, but straightforward computation, one finds:
\bea
 \rho&=&{k\,T_{D2}\,2\pi L_z\over 8 } V^{-1} \left(Z_3 -  K^2 \mathcal{F}_{z\theta} - {Z_3 K^1\over K^1+ V  \mathcal{F}_{z\theta}} \right)\nonumber\\
 &=& {\hat k\over 8 \pi q} V^{-1} \left(Z_3 - {4\pi q\, \rho_2\over \hat k} K^2 - \hat k V^{-1} {Z_3 K^1\over 4\pi q\,\rho_2 + \hat k {K^1\over V}} \right)\,,
\label{rhoreln1}
\eea
where, in the second line, we have used the relations \eqref{supertubecharges}, \eqref{tensions} and  \eqref{densities}, along with the fact that $\mathcal{F}_{z\theta}$ is independent of $z$. Using \eqref{radiusrelation}, (\ref{rhoreln1}) can be rewritten as:
\bea
 4\pi q ( 2\rho V + \rho_1 K^1 + \rho_2 K^2) = \hat k L_3 \,,
\label{bubble-DBI}
\eea
or again
\bea
  \hat k \rho =2\pi q\,\rho_1 \rho_2 \,.
\label{J-DBI}
\eea
Since the densities $\rho_1$, $\rho_2$ and $\rho$ are functions of $\psi$, these two equations are {\it functionnal equations}, depending on $\psi$. They exactly match \eqref{densrelna} and \eqref{densrelnc}. Hence the supergravity and the Born-Infeld calculation give the same regularity conditions for supertubes of arbitrary charge density.

%%%%%%%%%%%%%%%%%%%
\subsection{Obtaining the bubble equations from the DBI action}
%%%%%%%%%%%%%%%%%%%

In the last subsection, we showed that one can obtain the regularity conditions at the supertube location directly from the DBI action, even for a wiggling supertube.  Since the supertube is treated as a probe, the regularity conditions for the fully back-reacted solution at the other GH points - in other words the bubble equations - are not captured by the DBI analysis. Nevertheless, we will show here that the structure of the general solution outlined in Section \ref{sec:solution}, and in particular the regularity conditions at the GH points allows us to infer the complete set of functional bubble equations starting from the regularity condition at the supertube location. 

In order to obtain the bubble equations from the regularity conditions of a supertube with variable charge densities, we need to first understand these equations for a supertube with constant charge density in a multi-center GH solution. Since this solution does not
depend on the Gibbons-Hawking fiber, it descends to a multi-center four-dimensional solution of the type first constructed in \cite{Bates:2003vx}, where the GH centers are D6 and anti-D6 fluxed branes and the supertube is a fluxed D4 brane. For a solution with $N$ GH centers and one supertube, there are $N$ bubble or integrability equations insuring absence of Dirac-Misner strings at the GH points,
and one such equation for the supertube point. The sum of the $(N+1)$ bubble equations is zero.  

An important feature of these equations is that they only depend on the pairwise interaction between the centers, and do not contain terms that depend on the product of charges from three or more points. For example, each term in the supertube bubble equation depends on the supertube charges and on the charges at one of the GH centers. There is no term proportional to the product of two or more far-away GH points.

As we have shown in the explicit construction of the fully back-reacted solution, the two properties:  the fact that the bubble equations sum up to zero and only involve pairwise interactions, are still true for solutions with wiggling supertubes.  It is also easy to see that these two properties hold more generally for solutions with arbitrary number of centers and varying charge densities. Recall that the bubble equation at a given center can be obtained from the requirement that there be no Dirac-Misner (DM) string starting at that center.  Suppose one has already ensured the cancellation of DM strings at all points but one. Then the last point can only have a DM string ending at infinity and if one has already imposed regularity at infinity, it follows that the last bubble equation is automatically satisfied. This establishes the first property of the bubble equations, that is, their sum is zero. 

The pairwise property of the bubble equations follows from the general structure  of equation (\ref{omegaeqn}), determining the angular momentum vector $\vec \omega$: at both the GH points and at the supertube point, the right-hand side of equation (\ref{omegaeqn}) is sum of products of  the local charges with harmonic functions that satisfy {\it linear} equations. Hence, the contributions from various centers to the $\vec \omega$ equation for the supertube add linearly. 

In order to write the bubble equation for the supertube in the usual form, one should observe that equations (\ref{bubble-DBI}) and (\ref{J-DBI}) imply that if the supertube is round and hence it sits at a point in the $\IR^3$ base of the GH space, the two electric densities and the angular momentum density are not independent. Hence, this supertube has one function worth of degrees of freedom. 

If one now integrates equation (\ref{bubble-DBI}) over the supertube world-volume, one obtains the standard bubble equation for the supertube:
\be
{\hat Q}_1 K_1|_{ST} + {\hat Q}_2 K_2|_{ST}  - {\hat k} L_3|_{ST}+ J V|_{ST}=0\,,
\label{bubble-explicit}
\ee
where ${\hat Q}_1$ and ${\hat Q}_2$ are the total electric charges, and $J$ is the total angular momentum of the wiggly supertube (\ref{Jsuper}).

Hence, one can imagine taking a round supertube of given total charges ${\hat Q}_1$ and ${\hat Q}_2$ and letting the charge densities wiggle in accordance to (\ref{bubble-DBI})-\eqref{J-DBI}. The first three terms in the supertube bubble equation (\ref{bubble-explicit}) remain the same, and only the term proportional to the supertube angular momentum, $J$, is modified. Since one assumes the total charges to be fixed, it is clear that because there is only one independent density function,  the location of the tube will change in order to satisfy \eqref{bubble-explicit}.

If one now remembers that the $K_1$, $K_2$, $L_3$ and $V$ harmonic functions are sums of harmonic functions sourced at each of the Gibbons-Hawking points, one can easily see that the supertube bubble equation is entirely composed of terms of the form:
\be
{{\hat Q}_1 k_1^i  \over |\vec r_{\rm{ST}} - \vec r_i|}
\ee 
plus, possibly, some constants. 
 
Using the fact that all the terms in the bubble equations for the other GH center contain pairwise interaction terms, and the fact that the bubble equations sum up to zero, we find that the only effect of the wiggly supertube in the bubble equations for a given GH center, $i$, is to modify the terms 
\be
{J q^i + {\hat Q}_1 k_1^i + {\hat Q}_2 k_2^i - {\hat k} Q_3^i \over |\vec r_{\rm{ST}} - \vec r_i|}
\ee
by replacing $J,{\hat Q}_1,{\hat Q}_2$ and ${\hat k}$ with the corresponding quantities for the wiggly supertube. Furthermore, given that ${\hat k}$ does not oscillate, and that one is considering a supertube of constant total electric charges, the only modifications brought about by the wiggles show up in the expression of $J$.
 
Hence, the functional bubble equations for backgrounds containing a wiggly supertube in multi-center GH solutions are exactly the same as those for a round supertube, except that one replaces the angular momentum $J$ of the round supertube with the angular momentum of the wiggly one, which is obtained from the density modes via equations (\ref{Jsuper}), (\ref{J-DBI}) and (\ref{bubble-DBI}). 
  
%%%%%%%%%%%%%%%%%%%%%%%%%%%%%%%%%%%%%
\section{The entropy enhancement mechanism}
\label{sec:enhancement}
%%%%%%%%%%%%%%%%%%%%%%%%%%%%%%%%%%%%%

Having constructed the full solution corresponding to a wiggly back-reacted supertube in an ambipolar space, we now turn to the exploration of the amount of entropy such a supertube can store. In the regime of parameters where one can ignore the back-reaction of supertubes, this question was studied using the Born-Infeld action \cite{Bena:2008nh}, and it was found that the entropy a supertube can store depends on its location in the solution, and is much much larger than that of a supertube in flat space. In fact the entropy yielded by the Born-Infeld action {\it diverges} as the supertube approaches the hypersurface between the regions of opposite signature in the ambipolar GH base. It is therefore crucial to understand what are the limits of entropy enhancement, what cuts it off in the fully back-reacted solution, and what is the dependence on the total charges of the entropy that a supertube can store.

%%%%%%%%%%%%%%%%%%%
\subsection{A near-critical supertube}
%%%%%%%%%%%%%%%%%%%

The maximum entropy enhancement of a supertube potentially occurs when the supertube approaches the critical surface, $V=0$, or $r_+ = r_-$.
In this limit, $|\widehat F|$ (given by \eqref{hatFdefn}) is large. For generic values of the charges (such that $\sum_{A=1}^2 k_A \widehat Q_A \not= 0$), equation (\ref{bubbdiffb}) implies that the supertube angular momentum
$\widehat J$ becomes large in this limit, of order $|\widehat F|$. Moreover, the large $|\widehat F|$
limit of (\ref{Jhatres}) gives
\begin{equation}
 \widehat J ~=~  -\frac{ 2 \pi q \, k_1}{\hat k \, k_2} \int_0^{4\pi q}  \,    \rho_1(\psi)^2      \ d \psi  \,.
\label{largeF1}
\end{equation}
Hence, in $r_+ \to r_-$ limit both $\widehat J$ and the density fluctuation,  $\int_0^{4\pi q}  \,    \rho_1(\psi)^2      \ d \psi$, grow proportionally to $|\widehat F|$. In this limit, equation (\ref{densrelnd}) becomes
\be
\rho_2(\psi) = - {k_1\over k_2} \rho_1(\psi) + {4\pi q^2 \,k_1\over \hat k \, k_2^2} {\rho_1^2(\psi)\over \widehat F}\,;
\ee
note that the term proportional to ${\rho_1^2(\psi)\over \widehat F}$ remains finite in the limit, and hence it cannot be discarded.

%The first of these identities appears to imply that in this limit one has  $\sum_{A=1}^2 k_A
%\widehat Q_A = 0$, however this is not so, it simply means that in this limit the fluctuations are
%extremely large compared to the mean intrinsic electric charges,   $\widehat Q_A$, of the
%supertube.  We can therefore choose the $\widehat Q_A$ as we wish and in the limit
%$r_+ \to r_-$ the fluctuating densities must dominate the properties of the supertube.

We would like to argue that (\ref{largeF1}) means that  $\widehat J$ must be negative, and since we have taken the $k_I$ to be positive, this means that we need to argue that $\hat k$ is necessarily positive.  The simplest way to see this is to consider the scale factor of the Kaluza-Klein circle in the D1-D5-P duality frame:
\begin{equation}
W~\equiv~ {Z_3 \over \sqrt {Z_1 Z_2}}~=~ {K_1 K_2 + L_3 V \over\sqrt{(K_2 K_3+ L_1 V)(K_1 K_3 + L_2 V)}} \,.
\label{KKsize}
\end{equation}
On the critical surface, one has
\begin{equation}
W~=~   {\sqrt{K_1 K_2}   \over |K_3| } ~=~   \sqrt{k_1 k_2}   \bigg | k_3 +  { \hat k \,  r_\pm  \over 2\, \Delta} \bigg |^{-1}\,.
\label{critKKsize}
\end{equation}
If $\hat k$ and $k_3$ have opposite signs then there is a danger that the equation ${r_\pm\over \Delta} = -{2 k_3\over \hat k}$
will admit a solution and when this happens the warp factor $W$ diverges and the metric becomes singular.  We will therefore require that all the $k_I$ and $\hat k$ are positive.

%On the critical surface, $r_\pm$ is bounded below by $a$ and so if the supertube gets
%sufficiently close to the critical surface, the metric coefficient, $W$, is dominated by the $\hat k$ %term.  This would generate CTC's if $\hat k$ were negative, and so the supertube can only
%approach the critical surface if  $\hat k$ is positive. We will therefore require that all the $k_I$
%and $\hat k$ are positive.

This means that, in the near critical limit,  $\widehat J$ is necessarily negative and  proportional to the mean-square of the fluctuation density.  If the supertube has large density fluctuations then  $\widehat J$ is necessarily large and negative.

 %%%%%%%%%%%%%%%%%%%
\subsection{A simple example}
%%%%%%%%%%%%%%%%%%%

We now illustrate the results above by computing a very simple example. Take $\rho_1$ to be given by:
\begin{equation}
\rho_1(\psi)   ~=~  \frac{\widehat Q_1}{4 \pi q} \big( 1~+~ \alpha \cos \psi\big)\,,
\label{examplerho}
\end{equation}
that is, we only take the first Fourier mode. The other densities, $\rho_2$ and $\rho$, are then fixed by regularity as in (\ref{densrelnb}) and (\ref{densrelnc}). One can integrate (\ref{densrelnd}) to get
\begin{equation}
\widehat Q_2 ~=~ -\frac{\hat k k_1}{q} \,\widehat F ~+~   \frac{\hat k^2  k_1 k_2}{q^2} \,   \bigg[ \bigg(\widehat Q_1  ~+~\frac{\hat k k_2}{q} \widehat F \bigg)^2 ~-~ \alpha^2 \,\widehat Q_1^2  \, \bigg]^{-{1 \over 2}} \, \big(\widehat F^2 - 1\big)  \,,
\label{chgrelnb}
\end{equation}
which may be rewritten
\begin{equation}
\Big[ \widehat Q_2 ~+~  \frac{\hat k k_1}{q} \,\widehat F \Big] \, \bigg[ \bigg(\widehat Q_1  ~+~\frac{\hat k k_2}{q} \widehat F \bigg)^2 ~-~ \alpha^2\, \widehat Q_1^2  \, \bigg]^{ {1 \over 2}}~=~   \frac{\hat k^2  k_1 k_2}{q^2} \,  \big(\widehat F^2 - 1\big)  \,.
\label{chgrelnc}
\end{equation}
One can then get $\widehat J$ from (\ref{bubbdiffa})
\begin{equation}
 \widehat J  ~=~  -  \frac{ \hat k k_1 k_2 }{2 q^2}   ~-~  \frac{ 1 }{2 q} \, \bigg( \sum_{A=1}^2 k_A \widehat Q_A   \bigg)  \, \widehat F \,.
\label{chgrelnd}
\end{equation}

For fixed $\widehat Q_A$, (\ref{chgrelnc}) means that $\widehat F$, and hence, $r_+'/r_-'$ must be readjusted as the amplitude of the oscillation grows. For the sake of definiteness, assume that the $\widehat Q_A$ as well as the $k_j$ and $q, \hat k$ are all positive. In order to have a large entropy, we need to get a large, negative angular momentum. This is done by taking $r_+ - r_- \to 0_-$ so that $\widehat F$ is very large and positive. In this limit, one can see that $\alpha^2$ has to scale like $\widehat F$ in order to satisfy \eqref{chgrelnb}. One can thus assume in this limit, without loss of generality, that $1 <\! < \alpha  <\! < \widehat F$. Combining then (\ref{chgrelnc}) and  (\ref{chgrelnd}) one finds:
\begin{eqnarray}
 \widehat J  &=& \frac{ \widehat Q_1 \,  \widehat Q_2 }{2 \hat k } ~-~\alpha^2 \, \frac{ \widehat Q_1^2}{4 \hat k }  \Big[ \widehat Q_1 ~+~  \frac{\hat k k_2}{q} \,\widehat F \Big]^{-1}\, \Big[ \widehat Q_2 ~+~  \frac{\hat k k_1}{q} \,\widehat F \Big]    ~+~ {\cal O}\Big({ \alpha^4 \over \widehat F^2} \Big) \nonumber \\
  &=& \frac{ \widehat Q_1 \,  \widehat Q_2 }{2 \hat k } ~-~\alpha^2 \, \bigg[ \frac{ k_1}{4 \hat k k_2  }   \, \widehat Q_1^2    ~+~ {\cal O}\Big({ \alpha^2 \over \widehat F^2} \Big) \bigg]  \,.
\label{modangmom}
\end{eqnarray}
Thus taking $ \widehat J$ very large and negative corresponds to extremely large fluctuations, $\alpha$, and these can be sustained by taking the supertube extremely close to the critical surface ($\alpha^2\sim \widehat F$).

%%%%%%%%%%%%%%%%%%%
\subsection{Large-distance behaviour and an angular momentum bound}
%%%%%%%%%%%%%%%%%%%

If there is a large angular momentum then there is necessarily a limit on this given by the requirement that there are no CTC's at infinity. Since the supertube wraps the $\psi$ circles, we would expect the primarily limitation would come from this direction, and for fixed $t$ the metric coefficient of $(d \psi+A)^2$ is:
\begin{equation}
Y ~\equiv~  -(Z_1 Z_2 Z_3)^{-2/3} \, \mu^2 ~+~ (Z_1 Z_2 Z_3)^{1/3} \,V^{-1}  \,,
\label{psiscale}
\end{equation}
and this must be non-negative everywhere.

At infinity ($ r \to \infty$) one has:
\begin{equation}
Y ~\sim~  - { m_0 \over (k_1 k_2 (k_3 + {1 \over 2} \hat k ))^{1/3}}\, r  \,,
\label{CTCinf}
\end{equation}
from which it follows that $m_0$ must be negative.  It then follows from the bubble equations (\ref{bubbeqna}) that
\begin{equation}
 \frac{ k_1 k_2 k_3 }{q^2} \, \Big(\frac{r_+'}{a}  + \frac{r_-'}{a} \Big) ~+~    \frac{ \hat k k_1 k_2 }{2 q^2}  ~+~  \widehat J   ~=~ -\coeff{1}{2}\, m_0 (r_+' + r_-')~\ge~ 0\,.
\label{Jbound}
\end{equation}

This places a bound on how negative $\widehat J$ can be.  Indeed, if the supertube approaches the critical surface, then $\widehat J$, and hence the mean-square fluctuations, are limited, not by the intrinsic charges of the supertube, but by the product, $k_1 k_2 k_2$, and thus by the {\it total charge} of the corresponding black hole or black ring.

One can also use (\ref{bubbdiffb}) to write (\ref{Jbound}) as a bound on how close the supertube can come to the critical surface:
\begin{equation}
   \Big( \sum_{A=1}^2 k_A \widehat Q_A   \Big)\, \frac{a}{(r'_-  -  r'_+)}    ~\le~  \frac{ 2\,k_1 k_2 k_3 }{q}  \,.
\label{Fbound}
\end{equation}
Note that for the supertube to come extremely close to the critical surface the fluctuations must be large so that $\widehat J$ is large and negative, and so   (\ref{bubbdiffb}) means that $\widehat F$ should be large and positive, where one should remember that to arrive at this we have assumed that $r'_-  -  r'_+ >0$. It is important to stress that the DBI analysis, that only captures the {\it local} properties of the back-reacted solution, cannot encode large distances behavior, and hence the bound on $J$. The knowledge of the fully back-reacted solution was therefore crucial to obtain this bound.

%%%%%%%%%%%%%%%%%%%%%%%%%%%%%%%%%
\subsection{How to get most entropy from a supertube}
%%%%%%%%%%%%%%%%%%%%%%%%%%%%%%%%%

Recall that the entropy of the supertube is given by
\begin{equation}
S  ~\sim~ \sqrt{\widehat Q_1 \, \widehat Q_2  ~-~2\, \hat k  \widehat J } \,.
\label{Stubeentsimp}
\end{equation}
In flat space  $\hat k \widehat J $ is positive and proportional to the radius of the supertube and thus the angular momentum is bounded according to $| \widehat J|  \le |{1 \over 2 \hat k}  \widehat Q_1  \widehat Q_2|$, where the $\widehat Q_A$ are the intrinsic charges of the supertube.  On the other hand, it was shown in \cite{Bena:2008nh,Bena:2008dw}, using a brane-probe approximation, that in deep scaling solutions, the dipole-dipole interactions could allow $\hat k  \widehat J$ to become arbitrarily negative and therefore a supertube could store a vast amount of entropy, far beyond its flat-space limit.  This is the {\it entropy enhancement mechanism}.

We have now seen precisely the same process in the fully back-reacted solution, and how fluctuations in the deep-scaling limit of an AdS throat can indeed lead to very large negative values of $\hat k  \widehat J$ through dipole-dipole interactions as the supertube approaches the critical ($V=0$) surface.  The advantage of the back-reacted solution is that we can also see that there is a bound,  (\ref{Jbound}), on just how negative  $\widehat J$ can become.  If $\widehat F$ is large, one has $r_+' \approx r_-'$ and the supertube electric charges become irrelevant.  Thus the bound becomes:
\begin{equation}
- \widehat J ~\le~ \frac{ k_1 k_2 }{2\, q^2} \,(4\, \gamma\, k_3 ~+~ \hat k)   \,.
\label{simpJbound}
\end{equation}
where $\gamma \equiv {r_\pm' \over a}$ defines the aspect-ratio of the triangle defined by the supertube and the GH points.
If the flux parameters, $k_j$, are very large, then one has the following bound on the entropy enhancement:
\begin{equation}
 S  ~\sim~  \sqrt{\widehat Q_1 \, \widehat Q_2  ~-~2\, \hat k  \widehat J } ~\le~  \sqrt{\widehat Q_1 \, \widehat Q_2 +  \frac{\hat k  k_1 k_2 }{q^2} \,(4\, \gamma\, k_3 ~+~ \hat k) }   ~\sim~  \sqrt{\frac{\hat k  k_1 k_2 }{q^2} \,(4\, \gamma\, k_3 + \hat k)}  \,.
\label{Senhanced}
\end{equation}
Thus the entropy enhancement can be extremely large but is still limited.  In particular, the limit involves precisely the dipole-dipole interaction between the supertube and the background geometry.  In flat space the supertube entropy is limited by its intrinsic charges, whereas here it is limited by the {\it charges of the complete background}.  We have thus demonstrated that entropy enhancement is a very real, and potentially very large phenomenon but that it is bounded.

For one supertube, one can use the fact that the M2 charges of our solution are proportional to the product of the magnetic fluxes on the cycle between two GH centers to estimate the dependence of the entropy on the total charges. Assuming that the supertube dipole charge $\hat k$ is of the same order as the $k_i$, this gives $S \sim \sqrt{k^4} \sim \sqrt{Q^2}$. This entropy has the same growth with charges as that of a normal supertube in flat space; the difference is that it is not realized by putting all the charges on a supertube, but rather by using the charges to create a two-centered bubbled solution, and putting a very small supertube in this background.

Hence, the entropy of a single supertube in a bubbled background, though very enhanced, does not give a parametrically-larger entropy than that of a two-charge system. To get more entropy, we have to find a way to put a larger amount of negative angular momentum on this supertube, without destroying the asymptotics of the solution. It is not hard to see that the angular momentum bound (\ref{simpJbound}) is much like that of a BMPV black hole: the M2 charges of the background are proportional to the product of the fluxes: $Q_1 \sim k_2 k_3$, and the bound of the absolute value of $\hat J$ is
\be
|\hat J|~\le~ \sqrt{Q_1 Q_2 Q_3}~,
\ee
exactly like the BMPV black hole. However, it is well-known that for a given set of charges, the BMPV black hole is not the object with the largest angular momentum. A black ring, or a supertube can carry a parametrically-larger angular momentum, proportional to the square of the charges \cite{Mateos:2001qs,Bena:2004wt}.

Thus, a possible way to obtain more entropy than $\sqrt{Q^2}$ is to place more than one supertube in the two-center bubbling solution.  Indeed one could use a large non-wiggly supertube to act as an angular momentum sink and another smaller one that will give the entropy via the entropy enhancement mechanism. The two supertubes could be given dipole charges that are oriented in the same direction so that there is no danger of $W$ (equations (\ref{KKsize}) and  (\ref{critKKsize})) vanishing. However, the angular momentum of the non-wiggly supertube will point in the opposite direction from that of the entropy-enhanced supertube. Assuming that the dipole charge of the wiggly supertube is again of order $Q^{1/2}$, its angular momentum is now of order $Q^2$, and hence the total entropy of the system is
\be
S \sim \sqrt{\hat k |\hat J|} \sim \sqrt{Q^{5/2}}
\label{S-sink}
\ee
while this entropy is still not black-hole-like, it is parametrically larger than that of the two-charge system. Of course to establish that this kind of entropy is indeed present in our system, one would need to construct the fully-back-reacted solution with a wiggly and a non-wiggly supertube in global $AdS_3 \times S^2$  and to check that this metric is free of CTC's. 

Another possible way to obtain a larger entropy is to place many wiggly supertubes in this background. In general, the dipole charges of these supertubes should have the same orientation as the $k_3$ of the background, in order to avoid zeros in the scale factor $W$ in equation (\ref{critKKsize}). Hence, when they spin, their angular momenta will be oriented along the same direction, and will sum to the total allowable value of $|\hat J|$ (which is of the order $Q^{3/2}$). The total entropy will contain, besides the entropy of each individual supertube, a component coming from the many ways of partitioning $ \hat J$ between various supertubes, and will also be generically larger than $\sqrt{Q^2}$. Of course, by adding a large non-wiggly supertube that acts like an angular momentum sink we can increase the upper limit of $ |\hat J|$, and presumably obtain an entropy larger than even (\ref{S-sink}).

Probably the best way to bypass the angular momentum bound and obtain a large entropy is to put two counter-rotating wiggly supertubes, which naively could have huge opposite angular momenta, and could give us as much entropy as we want. However, there is no ``free lunch:'' the dipole moments of the two supertubes now have to be opposite, and for most of the possible locations of the supertubes, this will cause problems for the scale factor $W$ in equation (\ref{critKKsize}). However, it may be possible to select relative  locations for the two supertubes in the vicinity of the critical surface such that $W$ will never have a zero. If such a configuration exists, and satisfies the bubble equations, one can imagine starting to wiggle the supertubes, in order to get more and more states. As one does this, the bubble equations determining the relative positions of the points change, and one can imagine that at a certain upper value of the angular momenta these positions will become incompatible with an everywhere-positive $W$. It would be certainly interesting to explore this configuration in detail, and to see how much entropy can two oppositely-spinning supertubes store, and how does this entropy compare to that of a black hole with the same charges.

%%%%%%%%%%%%%%%%%%%%%%%%%%%%%%%%%%%%%
\section{Conclusions}
%%%%%%%%%%%%%%%%%%%%%%%%%%%%%%%%%%%%%

We have constructed an infinite-parameter family of smooth supergravity solutions that have three charges and three dipole charges, and that are microstates of black holes of classically-large horizon area. These solutions are obtained by placing a two-charge supertube of arbitrary shape in a two-center Gibbons-Hawking base space. In general the solutions for such supertubes can be implicitly written using scalar and vector Green functions, but so far no explicit solution has been written down, owing to the complicated form of the Green functions, and to the fact that for most of the base spaces that are physically of interest, the {\it ambipolar} GH spaces, the Green functions were unknown. 

In general, the oscillations of a supertube in the four-dimensional base space of the solution can be parameterized by four continuous functions but we have identified a subclass of supertube oscillations  where the shape of the supertube remains round, and only the distribution of electric charges inside the supertube world-volume changes. The solutions corresponding to these supertubes depend on one arbitrary continuous function, and their magnetic dipole fields are exactly the same as those of round supertubes, which considerably simplifies their explicit construction. The main ingredient that enters in the construction of the explicit solution is the scalar Green function on the base space. Given that the ambipolar GH base spaces have regions of signature $-4$ and $+4$, with intervening ``critical surfaces,''  the scalar Green functions is much more complicated than it is for regular GH spaces.  We could only find its explicit form for the two-center  ambipolar Gibbons-Hawking space given by the harmonic function ${1\over |\vec r + \vec a|} - {1\over |\vec r - \vec a|}$.  This was done via a highly non-trivial procedure that involved reducing the five-dimensional Green function on the smooth Lorentzian space-time that can be constructed from this base, that is,  from the Green function on global $AdS_3 \times S^2$.
 
In constructing this family of smooth horizonless black-hole microstate geometries, we have also found that the bubble (or integrability) equations that determine the relative locations of the Gibbons-Hawking centers and of the supertube are unchanged for the GH centers but become non-trivial functional bubble equations on the supertube. Remarkably, the same functional bubble equations can be recovered by examining the Born-Infeld action of a probe supertube in this space. Given the completely different nature of the two calculations that yield the same functional bubble equations, and given that one calculation is done in the regime of parameters where the supertube does not back-react on the geometry, while the other was done in the regime where it does, this result has several important implications.  First, it points to the existence of a non-renormalization theorem that protects the functional bubble equations as one moves in moduli space. Such a non-renormalization theorem exists for the four-dimensional multi-center solutions that come from a five-dimensional solution with a tri-holomorphic $U(1)$ invariance (and hence a Gibbons-Hawking base space), and our analysis finds that this non-renormalization extends to solutions that do not have this invariance. 

One of the most important uses of the non-renormalization theorem for these $U(1)$-invariant solutions has been the quantization and counting of certain finite-dimensional moduli spaces of multi-center configurations \cite{deBoer:2008zn,deBoer:2009un}. This is done at weak coupling, and then extrapolated to the regime of parameters where all the branes back-react using this non-renormalization theorem. The fact that the infinite-dimensional moduli space of wiggly supertubes does not receive quantum corrections implies that if one quantizes this moduli space at weak coupling (using for example the Born-Infeld action of supertubes) one can find how much entropy comes from fully-back-reacted wiggly supergravity solutions. 

Another important result of the agreement between the DBI analysis and the fully back-reacted descriptions of the bubbling solutions containing supertubes is that one can streamline the construction and analysis of supertube configurations that give smooth microstates without constructing the full supergravity solution for each and every microstate. Indeed, as shown in Section \ref{sec:fullsol} and in \cite{Bena:2008dw}, the local condition that the supergravity solution near the supertube be smooth and free of closed timelike curves is exactly the same as the condition that the supertube be a solution to the DBI action. In addition, we have also seen in this paper that the conditions that the solution be free of closed timelike curves near the GH centers, that is, all the other bubble equations,  can also be obtained from the functional bubble equations for the supertube.  Hence, if one wants to find the properties of a solution containing a supertube that has a charge density given by an arbitrary function $\rho_1(\theta)$, one first determines the density distribution for the other charge $\rho_2(\theta)$, using equation (\ref{densrelnb}), then finds the total supertube angular momentum density $J(\theta)$ using equation (\ref{densrelnc}). The next step is to integrate these densities to find the total charge and angular momentum, and to use the bubble equations to determine the location of this supertube and of the other Gibbons-Hawking centers in the full solution. Our analysis finds that as far as the full supergravity solution is concerned, the smooth supertube with variable charge and angular momentum density behaves exactly as a singular $U(1)$-invariant supertube that has the same total charges and angular momentum. Hence, to analyze whether a given wiggly supertube gives a smooth and regular geometry upon back-reacting one simply has to construct the fully back-reacted solution of an equivalent round supertube, which respects the tri-holomorphic $U(1)$ invariance of the GH base, and which is straightforward to write down in terms of harmonic functions. 

Last, but not least, our analysis finds that the entropy enhancement mechanism, first uncovered in \cite{Bena:2008nh} for non-back-reacted supertubes, extends to fully back-reacted solutions. This mechanism allows supertubes with relatively small electric charges to have a much larger entropy in a background with large magnetic fields than in flat space. We showed in Section \ref{sec:enhancement} that the entropy of such a supertube is bounded above not by its own electric charges, but by the electric charges of the background in which it is placed. This establishes that the entropy enhancement mechanism is not an artifact of the Born-Infeld approximation to the supertube dynamics, but is a feature of fully-back-reacted solutions containing supertubes.

Clearly, the most important open problem raised by our work is to determine how much entropy can be found in the class of solutions that we have constructed. In contrast to all three-charge solutions that have been obtained so far, whose moduli space is finite-dimensional, the solutions we construct have a much larger moduli space, whose dimension is infinite. Hence they should have a much larger entropy than that found by the semi-classical quantization of the highly symmetric bubbled solutions in  \cite{deBoer:2008zn,deBoer:2009un}. The question is how much more can be gained through the density fluctuations and through entropy enhancement. Indeed, the entropy of $U(1)$-invariant (and toroidally invariant) microstates solutions is parametrically smaller than that of the corresponding black holes, as one might have in hind-sight expected from counting only microstates that respect a certain isometry. It is also clear that the solutions we construct are not the most general black hole microstates one can finds in supergravity.  All of our solutions are independent of the internal space (which one can take to be $T^6$ or a more general Calabi-Yau), and utilize only one of the four functions worth of oscillations that a supertube can have in spacetime\footnote{For two-charge solutions that utilize the internal and fermionic degrees of freedom see \cite{Lunin:2002iz,Taylor:2005db,
Kanitscheider:2007wq}.}. Nevertheless, from the Born-Infeld action of supertubes we know that their moduli space is characterized by continuous functions, and the isometry we break does not destroy the nature of the moduli space and the nature of the counting problem, it just uses one of the eight bosonic degrees of freedom in the DBI action. Consequently, we expect the entropy of the solutions we construct to differ from the entropy of the supergravity solutions coming from all the possible supertube oscillations by a numerical factor of $\sqrt{8}$ (or $\sqrt{12}$ if one considers the fermion partners of the bosonic oscillations). Hence, counting the solutions we have constructed, while not capturing all the entropy of supergravity solutions, will give a finite and known fraction of the complete set of solutions. In contrast, the extra $U(1)$ isometry of the solutions counted in  \cite{deBoer:2008zn,deBoer:2009un} destroys the function-dependent nature of the moduli space and makes it finite-dimensional, hence such solutions cannot be expected to have an entropy that grows with the charges in the same way as that of the most general smooth supergravity solution.

On the more technical side, one drawback of our work is that we have not been able to obtain the expression of the four-dimensional rotation parameter $\vec \omega$ in closed form. The only danger associated with $\vec \omega$ is that it could lead to the appearance of closed timelike curves, either via Dirac-Misner strings, or more globally. The absence of the former  is guaranteed by the functional bubble equations. Furthermore, as we have argued above, all the global properties of a solution with smooth wiggly supertube can be captured by a Gibbons-Hawking solution that contains an equivalent round supertube. If the $\vec \omega$ of the latter solution does not cause closed timelike curves far-away from the supertube location, 
it  is likely that the $\vec \omega$ of the wiggly solution will not cause problems either. This being said, it would still be quite interesting to try to obtain $\vec \omega$ in closed form, at least for some of the solutions, and see that indeed the regularity of the solutions is insured by the functional bubble equation and by the regularity of the solution with an equivalent round supertube. Finding $\vec \omega$ explicitly would also probably be essential to the complete analysis of a solution with multiple counter-rotating, wiggling supertubes.  We are currently trying to develop a class of solutions that contain such supertubes.  

The work presented here has demonstrated the reality and viability of the entropy enhancement mechanism and thus represents significant progress in finding microstate geometries.  In subsequent work we hope to be able to construct solutions with multiple wiggling supertubes and show how they can interact with one another and generate mutual entropy enhancement while keeping the total angular momentum small.  The interactions and closed-timelike-curve analysis will limit the enhancement but  we remain hopeful that it will get us near the long-sought, semi-classical black-hole entropy.

%%%%%%%%%%%%%%%%%%%%%%%%%%%%%%%%%%%%%
\bigskip
\leftline{\bf Acknowledgements}
\smallskip
%%%%%%%%%%%%%%%%%%%%%%%%%%%%%%%%%%%%%
We would like to thank M. Berkooz, J. deBoer, A. Saxena, and S. El-Showk for valuable discussions.  NB, SG and NPW are grateful to the IPhT, CEA-Saclay for hospitality while much of this work was done. The work of IB, CR and SG was supported in part by the DSM CEA-Saclay, by the ANR grant 08-JCJC-0001-0, and by the ERC Starting Independent Researcher Grant 240210 - String-QCD-BH. The work of NB and NPW was supported in part by the DOE grant DE-FG03-84ER-40168.

%%%%%%%%%%%%%%%%%%%%%%%%%%%%%%%%%%%%%
\section*{Appendix A. Details of the Green function}
\appendix
\renewcommand{\theequation}{A.\arabic{equation}}
\setcounter{equation}{0} \addcontentsline{toc}{section}{Appendix
A. Details of the Green function}
 % reset counter
%%%%%%%%%%%%%%%%%%%%%%%%%%%%%%%%%%%%%

In the main body of the paper we only considered the details of the Green function, (\ref{Gnice}),  that were directly pertinent to finding the wiggling supertube solution.  In this Appendix we will examine the Green function in more detail and show that it does indeed have the requisite properties.

%%%%%%%%%%%%%%%%%%%%%%%%%%%%%%%%%%%%%
\subsection*{Appendix A1.  Branch cuts and singularities}
\appendix
%%%%%%%%%%%%%%%%%%%%%%%%%%%%%%%%%%%%%

We begin by considering the possible singularities of $\widehat G$ and while the physical domain has $|z| =1$, we will consider $|z| \le 1$ for the same reasons that were outlined in Section  \ref{Gprops}.  We will show that the only singularity is precisely the physical one exhibited in Section \ref{Gphyspole}.

From (\ref{params2}) it follows that $\sqrt{a_2 + i b_2} $ vanishes only if  $|u - v z|^2 = |y z |^2$ while from (\ref{params1}) one has $|u - v z| \ge  1$, $|y z| \le 1 $ with equality if and only if $z=1$, $\xi =\xi'$, $\theta = \theta' = \pi/2$.  Using this one then sees that $a_2 = b_2 =0$ if and only if  $\xi =\xi'$, $\psi=\psi'$, $\phi=\phi'$ and $\theta = \theta' = \pi/2$:  In other words both points coincide with each other and lie on the critical ($V=0$) surface.   Thus the only potential singularity generated by $\sqrt{a_2 + i b_2} $ occurs  only for a particular class of coincident points and this singularity was analyzed in \ref{Gphyspole}, and we will also discuss some further details below.   More generally, from the fact that $|u - v z| \ge  1$, $|y z| \le 1 $ it is not hard to convince oneself that $a_2$ is everywhere non-negative and so there is a globally analytic, single determination of the square root, which means that there are no branch points and branch cuts in the region of interest.

Now consider the singularities associated with the vanishing of the function, $\Lambda$.  From the last expression in (\ref{zLgeom})  we see that there are possible singularities in ${z \over \Lambda}$ when either  a) $\Delta =0$, b) $U = T =0$ or c)  $\theta = \pi - \theta'$. Note that while $|U|  \to \infty$ is certainly possible if $r_\pm + r_\pm' = \Delta$, the last  expression in (\ref{zLgeom})  shows that ${z \over \Lambda}$ remains finite.  The other expression in (\ref{zLgeom}), along with  (\ref{zmidents}), shows that there is, in fact no singularity associated with $\Delta$ vanishing by itself.   Similarly, the same expression, combined with the form of $m$ in (\ref{functions1}), shows that there is no singularity associated with $\theta = \pi - \theta'$ alone.   Thus the only singularity in (\ref{zLgeom}) occurs when  $U = T =0$, indeed there is a double pole if one sets $T=0$ first and then takes $U \to 0$.   We will now show that this double pole is generically cancelled by the other  terms in the complete Green function and that the only singularity arises when the points are coincident.

Setting $T=0$ in the Green function is equivalent to taking $z = {\bar m \over |m|}$.  In terms of the earlier notation, this means that:
\begin{equation}
 {  z \over \Lambda}  ~=~    {z_+  \over  m\, \big (z_+ -   {\bar m \over |m|} \big)^2}\,,
 \label{doublepole}
 \end{equation}
where we have used the fact that $z_+ z_-  =  {\bar m \over m}$.
The possible double pole thus emerges in the limit     $z_\pm  \to  {\bar m \over |m|}$.  Note the order of limits here:  The value of $z$ is fixed first and then the limit of $z_\pm$ is taken.  Also note that the choice $z = {\bar m \over |m|}$ has $|z|=1$ and is thus in the physical domain.  This means there is the possibility of  strongly singular behavior at non-coincident points in $G$ and we will now show that such singularities do not occur. 
 
We begin by assuming  $\Delta  \ne 0$ and hence  ${\cal P}   \ne 0$, then the limit of interest can only happen if $\sqrt{(1+x)^2 -y^2} = (\cos \theta + \cos \theta') =0 $.   Thus the potential singularity can only occur on an ``image surface'' at the ``conjugate latitudes,'' $\theta = \pi -\theta'$, on the $S^2$.

Let $\theta = \pi -\theta'- \varepsilon$ and so  $\sqrt{(1+x)^2 -y^2}  \sim  {\cal O}( \varepsilon)$. One   can also easily verify that
\begin{equation}
z ~=~  {\bar m \over |m|}  ~=~     {u + v\,e^{-i \chi}  \over u\,e^{-i \chi} + v } ~+~ {\cal O}( \varepsilon^2) \,, \qquad
z_\pm ~=~  {\bar m \over |m|} ~\pm~   {  \varepsilon  \over 2 m} \sin \theta \sqrt{{\cal P}} ~+~ {\cal O}( \varepsilon^2) \,, \label{mexp}
\end{equation}
and therefore
$(z_+ -   {\bar m \over |m|} )^2 \sim {\cal O}({\varepsilon^2})$.  On the other hand,
for physical parameter values ($|z|=1$) and non-coincident points, one has $ a_2 + i b_2 \ne 0$ but one also has
\begin{equation}
\nu ~=~   -(1+x) \big((u + v\, e^{-i \chi}  )  -  z \,(u\,e^{-i \chi} + v)\big)   ~+~ ((1+x) - y )(v -u z) e^{-i \chi}  ~\sim~ {\cal O}( \varepsilon^2) \,, \label{nuexp}
\end{equation}
where we have used (\ref{mexp}) and  $((1+x) - y) = \cos^2 {1\over 2} (\theta + \theta') \sim  {\cal O}( \varepsilon^2)$.  It follows that $\nu$ vanishes strongly enough to cancel the potential singularity as $\varepsilon \to 0$ and that the Green function (\ref{Gnice}) is, in fact, finite as $\varepsilon \to 0$.

Now suppose $\Delta =0$, then $\xi =\xi'$, $\theta = \theta'$ and $\phi = \phi'$, then $m = {r_+ r_- \over a^2}$ and $z_\pm =1$.  One then finds that:
\begin{equation}
 {  z \over \Lambda}  ~=~    {z  \over  m (z -   1)^2}\,,
 \label{properpole}
 \end{equation}
which indeed has a double pole as $z = e^{{i \over 2 q}(\psi -\psi')}  \to 1$.  This is, of course, the
correct double pole for the Green function as $\psi \to \psi'$ with $\xi =\xi'$, $\theta = \theta'$ and $\phi = \phi'$.

%%%%%%%%%%%%%%%%%%%%%%%%%%%%%%%%%%%%%
\subsection*{Appendix A2.  The scalar fields sourced by simple sources }
\appendix
%%%%%%%%%%%%%%%%%%%%%%%%%%%%%%%%%%%%%

In this appendix we compute the moments of the Green function by integrating it against the Fourier modes of a given electrical charge
distribution. By putting together these moments one can reconstruct  the full warp factor sourced by supertubes with variable charge distributions in the two-center ambipolar space discussed in this paper.

%%%%%%%%%%%%%%%%%%%
\subsubsection*{Fourier modes along the fiber}
%%%%%%%%%%%%%%%%%%%

Taking $q=1$ for the present, we compute the moments of the propagators, that is, the integrals:
\be
I_m = \int_0^{4\pi} \!\!\!d\psi \,e^{{i\,m\over 2}(\psi-\psi')} \, \widehat G={2\over i}\oint {dz\over z}\,z^m\, \widehat G\,.
\ee
where the $z$-integral is over a circle of radius 1.  We write the propagator as
\be
\widehat G={z\over 2\,a\, \Lambda} [\phi(z;\chi)+\phi(z^{-1};-\chi)]\,,\quad \phi(z;\chi) \equiv  {\nu\over  \sqrt{a_2+i b_2}}\,.
\ee
We assume $m\ge0$. The integrals with $m<0$ can be derived from the reality condition
\be
I_{-m}=I^*_m\,.
\ee
The function $\phi(z)$ is regular for $|z|<1$ but has a branch cut in the region $|z|>1$. Then, to compute the part of the integral containing $\phi(z^{-1})$ it is convenient to change the coordinate $z\to z^{-1}$.
\be
I_m={1\over i}\oint {dz\over a\, \Lambda}\,z^m\,\phi(z;\chi)+{1\over i}\oint {dz\over a\,\tilde\Lambda}\,z^{-m}\,\phi(z;-\chi)\,
\label{i12}
\ee
where
\be
\tilde\Lambda = z^2 \Lambda(z^{-1})= {\bar m} (z-z_+^{-1})(z-z_-^{-1})\,.
\ee
The first integral in (\ref{i12}) receives contributions only from the pole of $\Lambda$ inside $|z|<1$, which is $z_-$ if $U>0$ and $z_+$ if $U>0$:
\be
{1\over i}\oint {dz\over a\, \Lambda}\,z^m\,\phi(z;\chi)=2\pi {\cos\theta+\cos\theta'\over a \, m(z_+-z_-)}z_{\mp}^m\,,\quad \mathrm{if}\quad U\gtrless 0\,.
\ee
Using the identities derived above, one can re-write this integral in terms of $U$ and $T$:
\be
{\cos\theta+\cos\theta'\over a\, m(z_+-z_-)}={1\over \Delta}\,,\quad z_{\mp} = e^{\mp U} e^{-i \hat{T}}\,,
\ee
with
\be
e^{i \hat{T}} = e^{i T} z^{-1}\,,
\ee
and thus
\be
{1\over i}\oint {dz\over a\, \Lambda}\,z^m\,\phi(z;\chi)= {2\pi\over \Delta} e^{\mp m U} e^{-i m \hat{T}}\,,\quad \mathrm{if}\quad U\gtrless 0\,.
\ee
The second integral in (\ref{i12}) receives contributions both from the pole of $\tilde \Lambda$ (which is at $z=z_+^{-1}$ for $U>0$ and at  $z=z_-^{-1}$ for $U>0$) and from the pole of $z^{-m}$ at $z=0$. The residue at $z=z_{\pm}^{-1}$ is given by, for $U>0$
\bea
&&2\pi i {1\over a\, \bar m (z_+^{-1}-z_-^{-1})}\,\phi(z_+^{-1};-\chi) \,z_+^m = 2\pi i {1\over a\, m (z_- - z_+)}\,(-\phi(z_+);\chi)\,z_+^m\nonumber\\
&&=2\pi i {\cos\theta+\cos\theta'\over a\, m (z_+-z_-)}\,z_+^m= 2\pi i \,{1\over \Delta} e^{m U} e^{-i m \hat{T}} \,,
\eea
where we have used that
\be
\phi(z_+^{-1};-\chi)=-\phi(z_+;\chi)\,.
\ee
Analogously one finds the residue for $U<0$:
\be
2\pi i {\cos\theta+\cos\theta'\over a\, m (z_+-z_-)}\,z_-^m=2\pi i \,{1\over \Delta} e^{-m U} e^{-i m \hat{T}} \,.
\ee

The residue at $z=0$ is given by
\be
R_m\equiv{1\over (m-1)!} {d^{m-1}\over dz^{m-1}}\Bigl[{\phi(z;-\chi)\over a\, \tilde \Lambda}\Bigr]_{z=0}\,.
\ee
Note that
\be
R^*_m = {1\over (m-1)!} {d^{m-1}\over dz^{m-1}}\Bigl[{\phi(z;\chi)\over a\, \Lambda}\Bigr]_{z=0}\,.
\ee
We couldn't find a compact expression for $R_m$, nor re-write it in terms of $U$ and $T$.

Putting things together, we find
\be
{1\over 4\pi } I_m = {1\over \Delta} \cosh(m U)\,e^{-i m \hat{T}} + {R_m\over 2} \,.
\label{im}
\ee

%The canonically normalized propagator has $I_0={1\over 4\pi \Delta}$,
%thus we the canonically normalized propagator is
%\be
%{\widehat  G\over 16 \pi^2}\,.
%\ee
%We will continue to work with $\widehat  G$

Note that, contrary to the Page propagator, there is no cusp at $U=0$.

%%%%%%%%%%%%%%%%%%%
\subsubsection*{Some limits}
%%%%%%%%%%%%%%%%%%%

Let us first look at the limit in which $r'_+\to 0$. This is achieved by taking $\xi'=\epsilon/2$ and $\theta'=\epsilon$, and letting $\epsilon\to 0$, so that $r'_+\approx a \,\epsilon^2$. At $r'_+=0$ the $\psi'$ fiber degenerates: functions of $\psi'$ are not regular unless they appear in the combination
$\sqrt{r'_+} e^{\pm \psi'/2}\sim \epsilon \,  e^{\pm \psi'/2}$. Thus regularity requires that in this limit $I_m$ vanishes as
\be
I_m\sim \epsilon^{|m|}\,.
\ee
One can see that the two addends in (\ref{im}) separately diverge as
\be
{1\over \Delta} \cosh(m U)\,e^{-i m \hat{T}} \sim \epsilon^{-|m|}\,,\quad {R_m\over 2}\sim \epsilon^{-|m|}\,.
\ee
However one can check with Mathematica (we could do this only up to $m=2$) that the divergences cancel and that $I_m$ indeed vanishes as $\epsilon^{|m|}$.

Suppose that one of the points, say $x'$, is on the $V=0$ surface: this happens if $r'_+=r'_-$ or $\theta'={\pi\over 2}$. One then finds that $I_m$ goes to a finite value.

Finally, consider the locus $U=0$, or $r_+-r_-=-(r'_+-r'_-)$. This implies $\theta'=\pi-\theta$. Even in this limit one can check that $I_m$ is regular (and has no cusp).

%%%%%%%%%%%%%%%%%%%%%%%%%%%%%%%%%%%%
\bibliographystyle{utphys}
\bibliography{micro}

%%%%%%%%%%%%%%%%%%%%%%%%%%%%%%%%%%%%
\end{document}